\documentclass{sig-alternate-edit}
\usepackage{stmaryrd}
\usepackage{amssymb}
\usepackage{graphicx}
\usepackage{balance}
\usepackage[tight,footnotesize]{subfigure}
\usepackage{epsfig}
\usepackage[ruled,vlined]{algorithm2e}
\usepackage{algpseudocode}
\usepackage{amsmath}
\usepackage{caption}
\usepackage{times}
\usepackage{tabularx}
\usepackage{url}
\usepackage{color} 
\usepackage{hyphenat}
\usepackage{cases}
\usepackage{mdwlist}
\usepackage{cite}
\usepackage[table]{xcolor}
\usepackage[all]{nowidow}

\newtheorem{theorem}{Theorem}

\begin{document}


\title{Efficient and Accurate Path Cost Estimation \\Using Trajectory Data}
\author{
\alignauthor
Jian Dai$^{1,2}$\hspace{35pt} Bin Yang$^{2}$\hspace{35pt} Chenjuan Guo$^{2}$\hspace{35pt} Christian S. Jensen$^{2}$\\
\affaddr{$^1$Institute of Software, Chinese Academy of Sciences, Beijing, China}\\
\affaddr{$^2$Department of Computer Science, Aalborg University, Denmark}\\
\email{daijian@iscas.ac.cn\hspace{35pt}\{byang,cguo,csj\}@cs.aau.dk}
}

\maketitle

\begin{abstract}

Using the growing volumes of vehicle trajectory data, it becomes increasingly possible to capture time-varying and uncertain travel costs in a road network, including travel time and fuel consumption. The current paradigm represents a road network as a graph, assigns weights to the graph's edges by fragmenting trajectories into small pieces that fit the underlying edges, and then applies a routing algorithm to the resulting graph. We propose a new paradigm that targets more accurate and more efficient estimation of the costs of paths by associating weights with sub-paths in the road network. The paper provides a solution to a foundational problem in this paradigm, namely that of computing the time-varying cost distribution of a path.

The solution consists of several steps.
We first learn a set of random variables that capture the joint distributions of sub-paths that are covered by sufficient trajectories. Then, given a departure time and a path, we select an optimal subset of learned random variables such that the random variables' corresponding paths together cover the path. This enables accurate joint distribution estimation of the path, and by transferring the joint distribution into a marginal distribution, the travel cost distribution of the path is obtained.
The use of multiple learned random variables contends with data sparseness; the use of multi-dimensional histograms enables compact representation of arbitrary joint distributions that fully capture the travel cost dependencies among the edges in paths.
Empirical studies with substantial trajectory data from two different cities offer insight into the design properties of the proposed solution and suggest that the solution is effective in real-world settings.

\end{abstract}

\section{Introduction}

Due in part to the proliferation of networked sensors, notably GPS receivers built into smart-phones and other devices, increasing volumes of vehicle trajectory data are becoming available. A trajectory captures the travel of a vehicle in a road network and contains detailed information about the state of the roads it traverses at the time of travel.

We are also witnessing an increasing interest in understanding and taking into account the time-varying, detailed state of a road network.
It is of interest to know the travel times of paths throughout the day in order to plan when to travel and which path to follow, and also to calculate payments for transportation services in settings where such services are outsourced, e.g., in demand responsive transportation.
Further, as the transportation sector accounts for substantial greenhouse gas (GHG) emissions, it is of interest to know the time-varying GHG emissions of paths and thus enable eco-routing that reduces GHG emissions~\cite{ecosky}.

On this background, the natural question is how one can best utilize trajectory data to accurately and efficiently derive complex travel cost such as travel time or GHG emissions for any path in a road network at a given departure time.
This is the fundamental problem addressed in this paper. Solving this problem is important in its own right, and it also represents an important step towards enabling routing based on complex travel costs.

The conventional paradigm for solving the problem is to assign weights to edges and then to sum up the weights of a path's edges to compute the cost of the path~\cite{DBLP:conf/aaai/ZhengN13,DBLP:conf/aaai/IdeS11,DBLP:journals/tkde/YangKJ14,chen2005path,lim2013practical,DBLP:conf/icde/YangGJKS14}.
This conventional paradigm is inadequate in terms of accuracy and efficiency. The accuracy is adversely affected by the travel cost dependencies among different edges not being accounted for. In reality, the travel costs among edges can be dependent due to, e.g., turn costs.
The efficiency is adversely affected by the need to sum up as many weights as there are edges in a path. When having distributions as weights, the summing up (i.e., convolution~\cite{DBLP:conf/icde/YangGJKS14}) is expensive.

We propose a new paradigm that targets better accuracy and efficiency. In addition to assigning weights to edges, we assign weights to selected sub-paths. This leads to better accuracy because the weights of sub-paths better capture the travel cost dependencies among different edges in the sub-paths. This also enables improved efficiency because the number of expensive summing up operations are reduced.
In particular, the new paradigm addresses the following three challenges.

\begin{figure}[!htbp]
  \centering
  \subfigure[Complex Distribution]{
    \includegraphics[width=0.48\columnwidth]{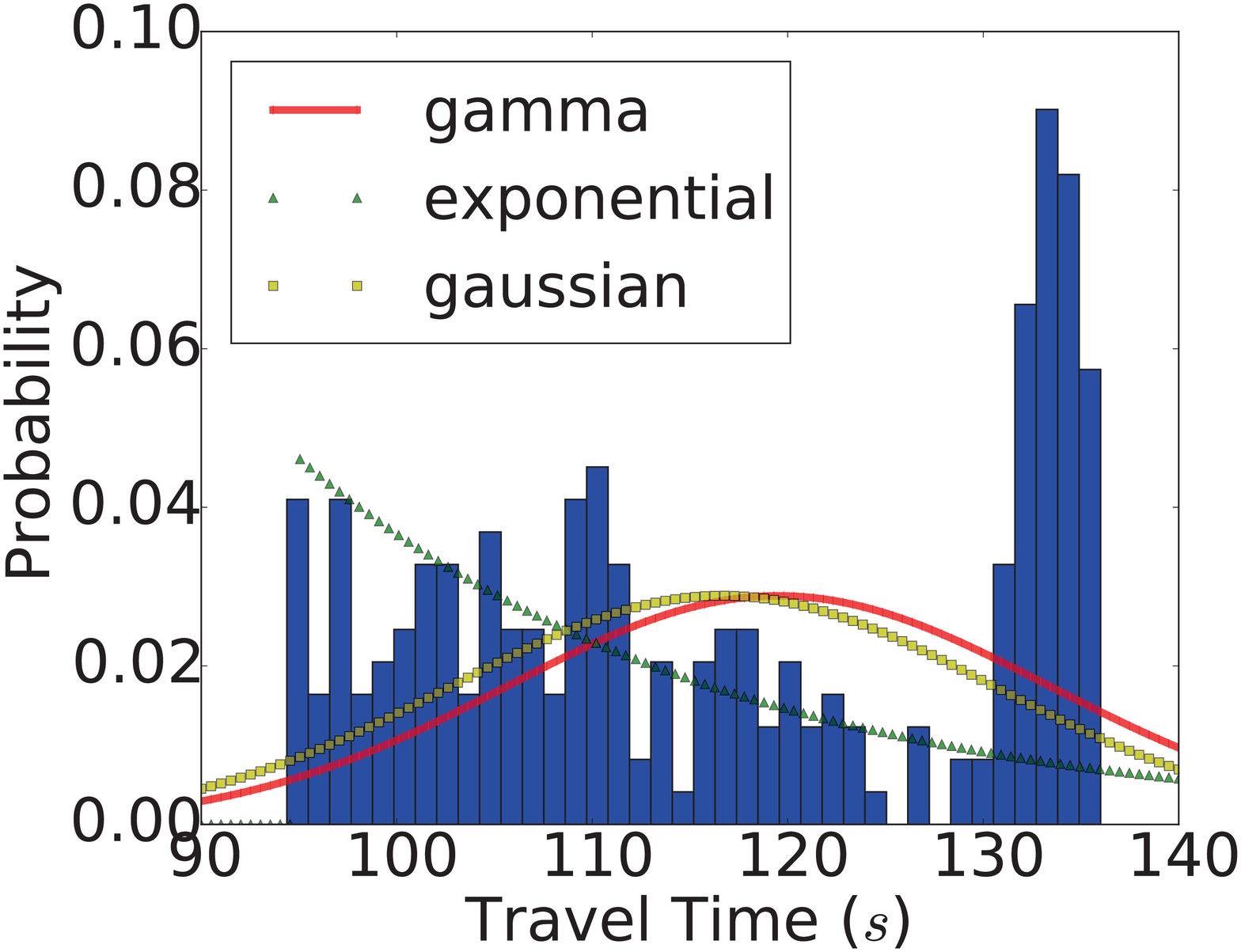}}
  \subfigure[$\mathcal{P}_1$ vs. $\mathcal{P}_2$]{
    \includegraphics[width=0.48\columnwidth,height=3.1cm]{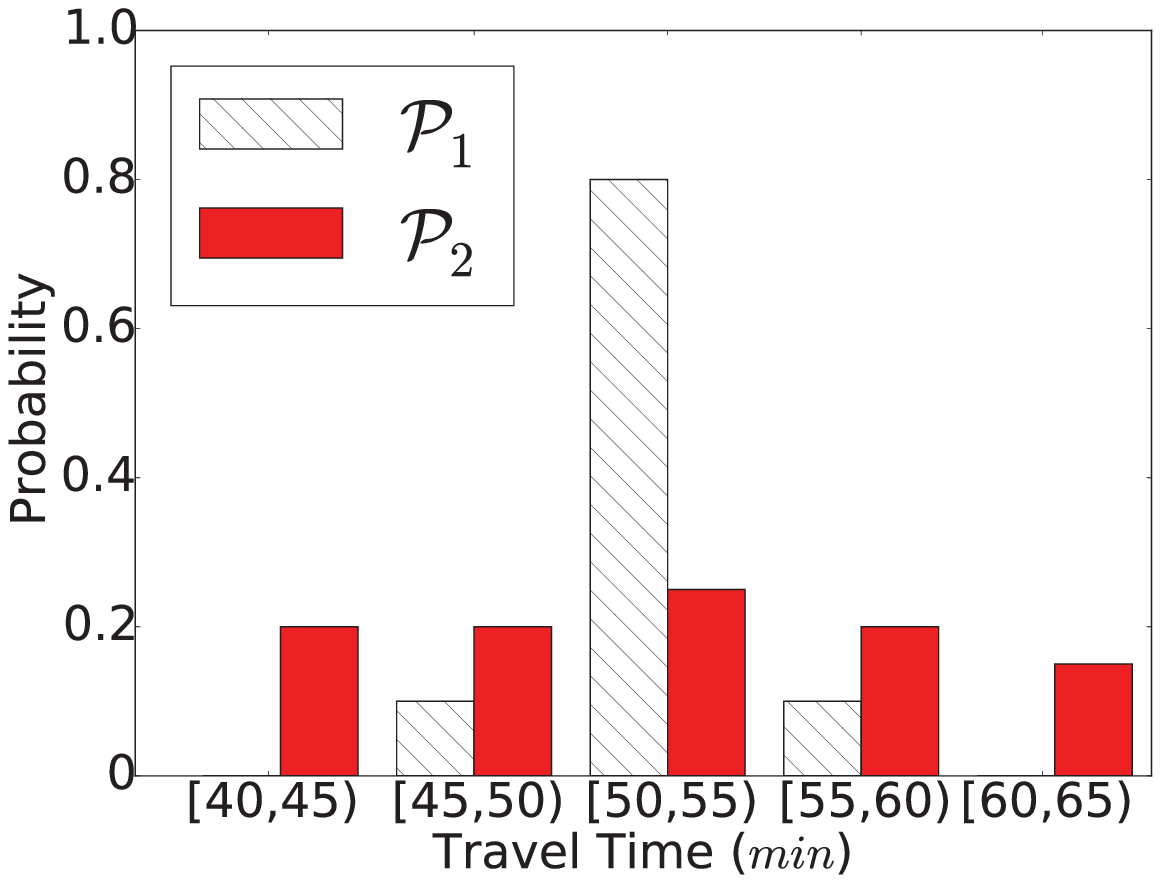}}
  \caption{Motivating Examples}
  \label{fig:introexample}
\end{figure}

\textit{Complex travel cost distributions}:
Unlike static travel costs, travel time and GHG emissions vary over time and even vary across vehicles traversing the same path at the same time. For example, the bars in Figure~\ref{fig:introexample}(a) represent the travel time, derived from GPS data, of a path during the time interval [8:00, 8:30). During other time intervals, the travel time is different.

Given a departure time and a path, we are interested in obtaining a cost distribution for traversal of the path when starting at the given time. The following example suggests why we aim for a cost distribution and not just a single cost value (e.g., a mean).
Assume that a person wants to drive from home to the airport. Path $\mathcal{P}_1$ uses highways, and path $\mathcal{P}_2$ goes through the city. Figure~\ref{fig:introexample}(b) shows the travel time distributions for the two paths at a given departure time.
If only considering single cost values, e.g., the means of the distributions, $\mathcal{P}_2$ (with mean 52 min) is always better than $\mathcal{P}_1$ (with means 52.5 min). However, $\mathcal{P}_1$ may be preferable if the person needs to arrive in the airport within 60 min to catch a flight because the probability of arriving at the airport within 60 min is 1, while with $\mathcal{P}_2$, the probability is 0.9. A 30-second average travel time penalty is a small price to pay for being sure to arrive on time. Thus, it is generally insufficient to simply report means; rather, distributions of costs are needed in many settings.

The accurate capture of time-varying aspects of travel calls for the modeling of path costs as time-varying distributions, where the distributions may be arbitrary, i.e., not following well-defined distributions.
For instance, the curves in Figure~\ref{fig:introexample}(a) show the corresponding Gaussian, Gamma, and exponential distributions that are obtained using maximum likelihood estimation. It is clear that the cost distribution does not follow standard distributions, e.g., Gaussian~\cite{DBLP:conf/aips/NikolovaBK06}, Gamma~\cite{DBLP:journals/corr/abs-1302-4987}, and exponential~\cite{DBLP:conf/uai/WellmanFL95} distributions. This also suggests that the travel cost estimation based on standard distributions is inaccurate in real-world settings.

\textit{Sparseness}: An attractive approach to deriving a time-varying distribution for a path is to use all trajectories that contain the path during the time interval of interest.
However, even with large volumes of trajectory data, it is difficult to cover all paths in a road network with sufficient numbers of trajectories during all time intervals because a road network has a very large number of meaningful paths. %
In general, the longer a path is, the more unlikely it is that it has sufficiently many trajectories to enable the derivation of an accurate distribution.

Thus, it is often the case that we have many short paths with accurate distributions. When deriving a long path's distribution, we need to sum up distributions of its shorter sub-paths.
However, it is challenging to efficiently sum up distributions. Given a long path, there may exist a large number of combinations of sub-paths. Considering all possible combinations is prohibitively expensive.

\textit{Dependency}:
When deriving the distribution for a path from distributions of sub-paths, these distributions are generally not independent. Two existing studies~\cite{chen2005path,lim2013practical} derive distributions of paths from distributions associated with their edges while assuming independence, and one study~\cite{DBLP:conf/icde/YangGJKS14} assumes that edges have time-varying distributions that are independent conditioned on the arrival times at the edges. We offer evidence that these independency assumptions generally do not hold and thus yield inaccurate cost distribution of the long path.
To derive accurate travel cost distributions for paths, the dependencies among different edges in a path, which are reflected in the trajectories,  must be considered.

We propose a novel paradigm that generalizes the convention paradigm. Within this paradigm, we provide techniques that contend with the three above challenges and that are able to derive accurate travel-cost distributions of any path in a road network.
Based on a large trajectory set, we learn a set of random variables that capture the joint distributions of paths that are covered by sufficient amounts of trajectories.
To contend with \textit{complex} cost distributions, an adaptive multi-dimensional histogram approach is proposed to approximate arbitrary distributions in an accurate and compact manner.

Given a departure time and a path, we select an optimal subset of the variables such that the paths corresponding to the variables together cover the query path at the departure time and such that the set of variables produces the most accurate joint distribution of the edges in the path. The joint distribution is then transferred into a marginal distribution that captures the travel cost distribution of the path.
We prove the property that an optimal subset should have and then propose a method to efficiently identify the optimal subset. The optimality ensures the accuracy of the computed travel cost distribution of the given path.  %
Deriving the distributions of longer paths with insufficient trajectories from the distributions of carefully selected sub-paths with sufficient trajectories solves the \textit{sparseness} problem.
The joint distribution of a path fully captures the dependencies among the edges in the path, thus addressing the \textit{dependency} problem.

We believe that this is the first study to enable accurate travel cost distribution estimation of a path using trajectories while contending with sparseness, dependency, and distribution complexity. The study provides evidence that when costs are derived from trajectory data, the conventional paradigm of first assigning weights or distributions to edges is inferior to the more general paradigm where weights are associated with  sub-paths.

To summarize, the paper makes four contributions. First, it proposes techniques to identify a set of learned random variables that captures the joint distributions of frequently-traversed paths. Second, it proposes an algorithm that identifies an optimal subset of learned random variables, enabling accurate estimation of the joint distribution of a query path. Third, it provides a method that derives the marginal distribution, represented as a one-dimensional histogram, from a joint distribution, represented as a multi-dimensional histogram.
Fourth, extensive empirical studies on two large trajectory sets are offered to elicit the design properties of the paper's proposal.

\emph{Paper outline}: Section~\ref{sec:pre} covers basic concepts and baselines. Section~\ref{sec:costDep} describes the identification of learned random variables. Section~\ref{sec:TCDE} gives the algorithms for estimating the travel cost of a path. Section~\ref{sec:exp} reports on the empirical study. Related work is covered in Section~\ref{sec:rw}. Conclusions  are offered in Section~\ref{sec:conclusion}.

\section{Preliminaries}
\label{sec:pre}

We introduce basic concepts and cover baseline algorithms.

\subsection{Basic Concepts}
\label{ssec:basiccp}

A \textit{road network} is modeled as a directed graph $G=(V, E)$,
where $V$ is a vertex set and $E \subseteq V \times V$ is an edge
set.
A vertex $v_i \in V$ represents a road intersection or an end of a road.
An edge $e_k=(v_i, v_j) \in E$ models a directed road segment,
indicating that travel is possible from its \emph{start vertex} $v_i$ to its
\emph{end vertex} $v_j$.
We use $e_k.s$ and $e_k.d$ to denote
the start and end vertices of edge $e_k$.
Two edges are \emph{adjacent} if one edge's end vertex is the same as the other edge's start vertex.

A \textit{path} $\mathcal{P}=\langle e_1, e_2, \ldots, e_{a}
\rangle$, $a \geqslant 1$, is a sequence of adjacent edges that connect distinct vertices in the graph, where $e_i\in
E$, $e_i.d=e_{i+1}.s$ for $1 \leqslant i < a$, and the vertices $e_1.s$, $e_2.s$, $\ldots$, $e_a.s$,
and $e_a.d$ are distinct.
The cardinality of path $\mathcal{P}$, denoted as $|\mathcal{P}|$, is
the number of edges in the path.
Path $\mathcal{P}'=\langle g_1, g_2, \ldots, g_{x}\rangle$ is a \textit{sub-path} of $\mathcal{P}=\langle e_1, e_2, \ldots, e_{a}\rangle$ if $|\mathcal{P}'| \leqslant |\mathcal{P}|$ and there exists an edge $e_i \in \mathcal{P}$ such that $g_1=e_i$, $g_2=e_{i+1}$, $\ldots$, and $g_x=e_{i+x-1}$.

Given two paths $\mathcal{P}_1=\langle e_1$, $e_2$, $\ldots, e_{a} \rangle$ and $\mathcal{P}_2=\langle g_1, g_2, \ldots, g_{b}\rangle$, the \textit{concatenation} of $\mathcal{P}_1$ and $\mathcal{P}_2$, denoted as $\mathcal{P}_1 \circ \mathcal{P}_2$, is defined as $\langle e_1, e_2$, $\ldots$, $e_{a}, g_1, g_2$, $\ldots$, $g_{b}\rangle$ if edges $e_a$ and $g_1$ are adjacent, i.e., $e_a.d=g_1.s$. Otherwise, the concatenation is an empty sequence.

A \textit{trajectory} $\mathcal{T}=\langle p_1, p_2, \ldots, p_b
\rangle$ is a sequence of GPS records pertaining to a trip, where each
 $p_i$ is a $(\mathit{location}, \mathit{time})$ pair of
a vehicle, where $p_{i}.\mathit{time}<p_{j}.\mathit{time}$ if
$1\leqslant i<j\leqslant b$.

Map matching~\cite{DBLP:conf/gis/NewsonK09} is used to map a GPS record to a
specific road network location and is able to transform a trajectory $\mathcal{T}$ into a sequence of
\textit{edge records} $\langle l_1, l_2, \ldots, l_C \rangle$.
A record $l_i$ is of the form $(e, t, \mathit{GPS})$, where $e$ is
an edge traversed by trajectory $\mathcal{T}$; $t$ is the time when
the traversal of edge $e$ starts; and $\mathit{GPS}=\langle p_j$, $p_{j+1}$,
$\ldots$, $p_k \rangle$ contains the GPS records mapped to edge $e$.

The sequence of the edges in the sequence of edge records is referred
as the \textit{path of trajectory} $\mathcal{T}$, denoted as
$\mathcal{P}_{\mathcal{T}}=\langle l_1.e, l_2.e$, $\ldots$, $l_C.e
\rangle$.
A trajectory $\mathcal{T}$ \textit{occurred on} path $\mathcal{P}$ \textit{at} $t$ if and only if path $\mathcal{P}$ is a sub-path of the path of trajectory $\mathcal{P}_{\mathcal{T}}$ and the first GPS record in the first edge in path $\mathcal{P}$ is obtained at time $t$.

The travel cost (e.g., travel time or GHG emissions) of using a path $\mathcal{P}$ can be obtained from the trajectories that occurred on $\mathcal{P}$. Given a trajectory $\mathcal{T}$ that occurred on path $\mathcal{P}$ at $t$, the travel time of using $\mathcal{P}$ at $t$ is the difference between the time of the last GPS record and the time of the first GPS record on path $\mathcal{P}$; and the GHG emissions of using $\mathcal{P}$ at $t$ can be computed from the edge records on path $\mathcal{P}$ using vehicular environmental impact models~\cite{DBLP:conf/gis/GuoM0JK12,ecomark2}.

\subsection{Problem Definition}

Given a large collection of trajectories $\mathbb{T}$ that occurred on a road network $G$, \emph{travel cost distribution estimation}, $TCDE(\mathcal{P}, t)$, takes as input a path $\mathcal{P}$ in $G$ and a departure time $t$, and \emph{accurately} and \emph{efficiently} estimates the travel cost of traversing path $\mathcal{P}$ at $t$.
Specifically, the output of $TCDE(\mathcal{P}, t)$ is a univariate random variable that describes the distribution of the total travel cost of traversing path $\mathcal{P}$ at $t$. The travel cost distribution is derived from the travel costs of the trajectories that occurred on path $\mathcal{P}$ or on sub-paths of $\mathcal{P}$.

The travel cost distribution estimation aims to be efficient to support interactive use. Notably,  travel cost distribution estimation is a fundamental operation in stochastic routing algorithms. Since such algorithms typically require efficient evaluation of the cost distributions of multiple candidate paths, efficient travel cost distribution estimation is essential for the efficiency of routing algorithms.

\subsection{Baseline Algorithms}
\label{ssec:aba}
\textbf{Accuracy-Optimal Baseline:}
An accurate way to estimate the travel cost distribution of path $\mathcal{P}$ at time $t$ is to retrieve a set of qualified trajectories. A trajectory $\mathcal{T}$ is qualified if $\mathcal{T}$ occurred on $\mathcal{P}$ at $t'$ and the difference between $t'$ and $t$ is less than a threshold, e.g., 30 minutes.

For each qualified trajectory, we compute a travel cost value for traversing path $\mathcal{P}$ at time $t$ (cf. Section~\ref{ssec:basiccp}).
A qualified trajectory captures the traffic conditions (e.g., the time it takes to pass intersections, wait at traffic lights, and make turns at intersections) of the entire path $\mathcal{P}$ during the interval of interest. Thus, no explicitly modeling of complex traffic conditions at intersections is needed.
And the travel cost value obtained from a qualified trajectory is very accurate. When having sufficient amounts of qualified trajectories, we are able to accurately derive the travel cost distribution of path $\mathcal{P}$ at $t$ based on the travel cost values derived from the qualified trajectories.
We regard the distribution $\mathbf{D}_{\mathit{AOB}}(\mathcal{P}, t)$ obtained by this baseline as the \emph{ground truth} distribution.

\begin{figure}[!hpt]
  \centering
  \subfigure[Aalborg]{
    \label{fig:subfig:spatial} 
    \includegraphics[width=0.48\columnwidth]{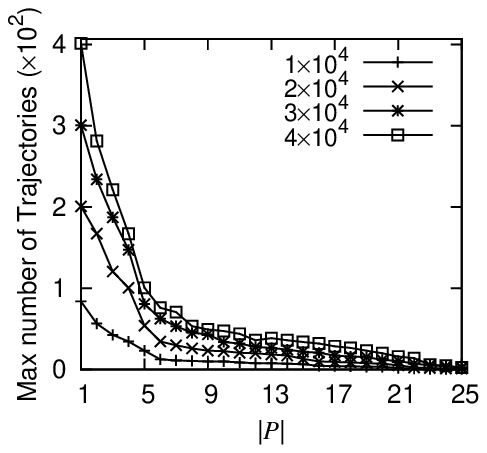}
    }
  \subfigure[Beijing]{
    \label{fig:subfig:temproal} 
    \includegraphics[width=0.48\columnwidth]{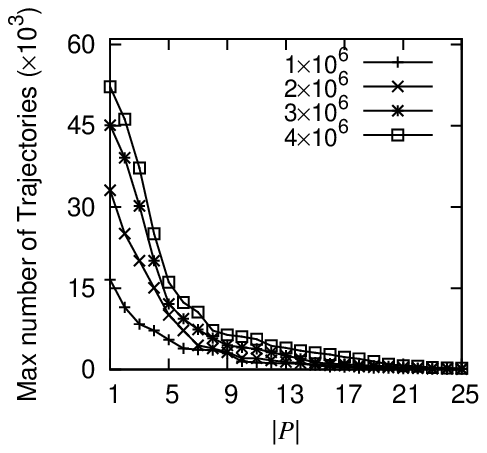}}
  \caption{Data Sparseness Problem}
  \label{fig:data-sparsity} 
\end{figure}

However, the accuracy-optimal baseline is often inapplicable due to \emph{data sparseness}: the cardinality of the set of qualified trajectories is often very small or even zero, especially for long paths.
Figure~\ref{fig:data-sparsity} shows that the maximum number of trajectories that occurred on a path decreases rapidly as the cardinality of the path increases, based on two large collections of trajectories in Aalborg and Beijing.
Figure~\ref{fig:data-sparsity} also suggests that as the number of trajectories increases, the maximum number of trajectories that occurred on a path with large cardinality (e.g., larger than 10) increases only little.

The data sparseness problem renders the accuracy-optimal baseline inapplicable in many cases, and we need a method that is able to accurately estimate the travel cost distribution of any path using \emph{sparse} trajectories.

\textbf{Legacy Baseline:}
The legacy baseline, which is extensively used in the conventional paradigm, operates at edge granularity and makes the \emph{independence assumption}~\cite{DBLP:conf/aips/NikolovaBK06,lim2013practical,chen2005path,DBLP:conf/icde/YangGJKS14}
that the travel cost distributions of the edges in a path are independent of each other.
Thus, a path's travel cost distribution is the \emph{convolution} of the travel costs distributions of the edges in the path.
In most existing studies based on the independence assumption~\cite{DBLP:conf/aips/NikolovaBK06,lim2013practical,chen2005path}, the travel cost distributions of edges are synthetically generated and are not derived from real-world traffic data.

We follow a recent study~\cite{DBLP:conf/icde/YangGJKS14} that employs the legacy baseline.
Path $\mathcal{P}$ is partitioned into $|\mathcal{P}|$ sub-paths, where each sub-path consists of one edge.
For each sub-path, we use the accuracy-optimal baseline to retrieve a set of qualified trajectories and  use these to compute a distribution based on the travel cost values.
Since each sub-path has one edge, it is very unlikely that we have the sparseness problem when using the accuracy-optimal baseline.

The distributions from all sub-paths are summed into a distribution for path $\mathcal{P}$.
Since we assume that the distributions on edges are independent of each other, the sum of the distributions is then given as the convolution of the distributions from all sub-paths in the path.
We denote the distribution obtained by the legacy baseline as $\mathbf{D}_{\mathit{LB}}(\mathcal{P}, t)$, which is defined in Equation~\ref{eq:conv}.
\begin{equation}
\label{eq:conv}
\mathbf{D}_{\mathit{LB}}(\mathcal{P}, t)=\bigodot_{e_i\in \mathcal{P}} \mathbf{D}_{\mathit{AOB}}(\langle e_i\rangle, I_{\langle e_i\rangle}),
\end{equation}
where $\bigodot$ denotes the convolution between two distributions and $\mathbf{D}_{\mathit{AOB}}(\langle e_i\rangle, I_{\langle e_i\rangle})$ denotes the travel cost distribution of sub-path $\langle e_i\rangle$ using the accuracy-optimal baseline, where $I_{\langle e_i\rangle}$ is the arrival time on edge $e_i$, which may be different from the departure time $t$.
Specifically, for the first edge $e_1$, we have $I_{\langle e_1\rangle}=t$. For an edge $e_i$, where $i>1$,  $I_{\langle e_i\rangle}$ needs to be progressively updated according to the travel times of $e_i$'s predecessor edges~\cite{DBLP:conf/icde/YangGJKS14}.

However, the independency assumption does not always hold---the travel cost distributions on two adjacent edges may be highly dependent, which then renders the convolution-based distribution inaccurate.
To demonstrate this, we identify 2,000 paths that each consists of two adjacent edges and on which at least 100 trajectories occurred within interval [7:30, 8:00) from GPS data.

\begin{figure}[!htbp]
\centering
\begin{tabular}{@{}c@{}c@{}}
    \includegraphics[width=0.48\columnwidth]{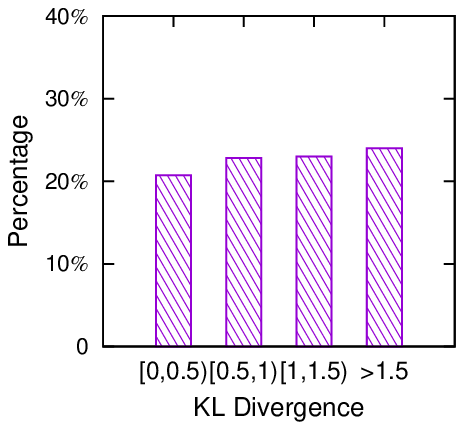}
    &
    \includegraphics[width=0.48\columnwidth]{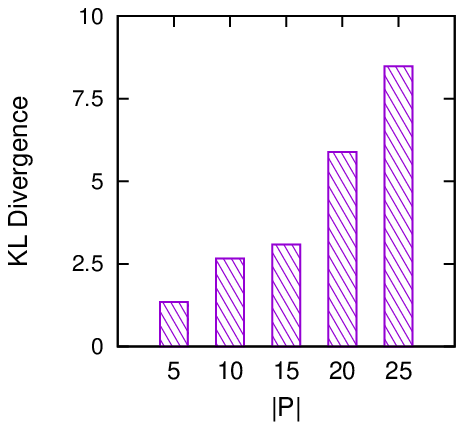}
    \\
    \small(a) $KL(D_{\mathit{AOB}}, D_{\mathit{LB}})$
    &
    \small(b) Varying $|\mathcal{P}|$
\end{tabular}
\caption{Examining the Independence Assumption} \label{fig:pairKL}
\end{figure}

For each path $\mathcal{P}$, we compute the distributions $D_{\mathit{AOB}}$ and $D_{\mathit{LB}}$ using the accuracy-optimal baseline and the legacy baseline, respectively. Since each path has at least 100 trajectories, the accuracy-optimal baseline faces no sparseness problem.
If the travel cost distributions of the two edges of a path are independent, the two distributions $D_{\mathit{AOB}}$ and $D_{\mathit{LB}}$ should be identical. To see if this holds, we compute the KL-divergence of $D_{\mathit{LB}}$ from $D_{\mathit{AOB}}$, denoted as $\mathit{KL}(D_{\mathit{AOB}}, D_{\mathit{LB}})$, which indicates how different the convoluted distribution $D_{\mathit{LB}}$ is from the ground-truth distribution $D_{\mathit{AOB}}$. The larger the KL-divergence, the more different the two distributions are, meaning that the convoluted distribution is less accurate.
Figure~\ref{fig:pairKL}(a) suggests that most of the adjacent edges are not independent, i.e., have large KL-divergence values.

Next, we conduct an experiment on 100 paths with different cardinalities (i.e., number of edges) and where each path has at least $30$ qualifying trajectories during an interval.
After computing $D_{\mathit{AOB}}$ and $D_{\mathit{LB}}$ for each path, the average KL-divergence values between the two distributions for the paths with certain cardinalities are shown in Figure~\ref{fig:pairKL}(b). This suggests that the more edges a path has, the more different the convoluted distribution $D_{\mathit{LB}}$ is from the true distribution $D_{\mathit{AOB}}$. Hence, the legacy baseline method is likely to give inaccurate travel cost distributions, especially for long paths.
This is not surprising because the independency assumption ignores or does not fully capture complex factors such as  turn costs at intersections, which are captured by the accuracy-optimal baseline method.

Further, as convolution is an expensive computation, the legacy baseline method is inefficient. The longer a path is, the more convolutions are needed.
An efficient method that is able to fully capture the dependencies among edges in a path is necessary.

\subsection{Solution Overview}

The analyses of the baselines suggest that when we compute the cost distribution for a path, we should try to use trajectories that occurred on long sub-paths of the path because they capture many hard-to-formalize factors such as turn costs. The analyses also suggest that the independency assumption does not always hold and that explicitly modeling of the dependency among the travel costs of different edges in a path is needed to achieve accurate results.

\begin{figure}[!htb]
  \centering
  \includegraphics[width=\columnwidth]{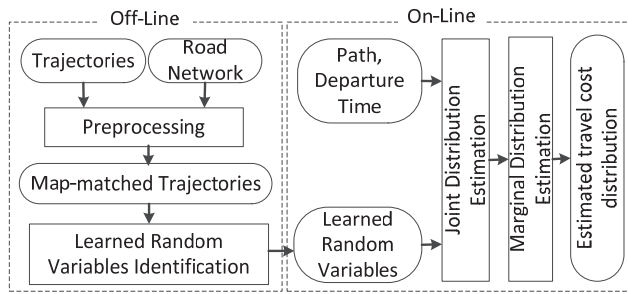}
  \caption{Solution Overview}
  \label{fig:sys-arch}
\end{figure}

An overview of the new paradigm is provided in Figure~\ref{fig:sys-arch}. In the off-line component, a pre-processing module takes trajectories and a road network as input and performs map matching. The map-matched trajectories are then used for generating a set of learned random variables, where each describes the joint travel cost distribution of a path during an interval. The paths in the learned random variables should have sufficient numbers of qualifying trajectories, and thus are generally of low cardinality.

The online component takes a path and a departure time as query arguments. The \emph{joint distribution estimation} module chooses an optimal sub-set of learned random variables such that their paths together cover the query path; and it accurately estimates the joint distribution of the query path based on the chosen variables. Next, the \emph{marginal distribution estimation} module transfers the estimated joint distribution of the path into a marginal distribution of the path, which is the distribution of the travel cost of using the path at the given departure time.

\section{Modeling Cost Dependency}
\label{sec:costDep}

As Figure~\ref{fig:pairKL} suggests, the travel costs of edges exhibit varying degrees of dependency.
To model the dependency among the edges in a path, we derive joint distributions of the edges' travel costs from the trajectories that occurred on the path.
We describe how to derive the joint distributions of multiple edges in paths.
The procedure is conducted \emph{off-line}.

\subsection{Deriving Distributions For Unit Paths}
\label{ssec:distunitpath}

A \emph{unit path} consists of a single edge.
The travel cost distributions on a unit path may differ significantly during different time intervals due to, e.g., peak vs. off-peak traffic. Thus, it is of interest to derive time-dependent distributions for unit paths.

\subsubsection{General Procedure}

We partition a day into $\lambda=\lceil\frac{24\times 60}{\alpha}\rceil$ intervals, where parameter $\alpha$ specifies the finest-granularity interval of interest in minutes, e.g., 30 minutes.
We let $V_{\langle e_i \rangle}^{I_j}=p(c_{ e_i  })$ denote a random variable that describes the travel cost distribution on unit path $\langle e_i \rangle$ during interval $I_j$.
The travel cost distribution $p(c_{ e_i })$ on unit path $\langle e_i \rangle$ is a distribution of a single variable $c_{e_i}$ representing the cost of traversing $\langle e_i \rangle$.

To derive the distribution of $V_{\langle e_i \rangle}^{I_j}$, a qualified trajectory set $\mathbb{T}_{\langle e_i \rangle}^{I_j}$ of trajectories that occurred on $\langle e_i \rangle$ at $t$ where $t \in I_j$ is obtained.

If the cardinality of the trajectory set $\mathbb{T}_{\langle e_i \rangle}^{I_j}$ exceeds a threshold $\beta$, the travel cost values obtained from the trajectories in $\mathbb{T}_{\langle e_i \rangle}^{I_j}$ are employed to instantiate the distribution of $V_{\langle e_i \rangle}^{I_j}$. The details of the instantiation are covered in Section~\ref{sssec:repuni}.

If the cardinality of $\mathbb{T}_{\langle e_i \rangle}^{I_j}$ does not exceed $\beta$, the distribution of $V_{\langle e_i \rangle}^{I_j}$ is derived from the speed limit of edge $e_i$ to avoid overfitting to the limited number of travel costs in $\mathbb{T}_{\langle e_i \rangle}^{I_j}$.

Thus, for one day, each unit path is associated with $\lambda$ random variables and corresponding qualified trajectory sets.

Next, if two random variables from two consecutive time intervals are similar, the corresponding qualified trajectory sets are combined, and the intervals are merged into a longer interval. A new random variable representing the travel cost distribution for the long interval is instantiated from the combined qualified trajectories.
The process continues until no random variables from consecutive intervals can be combined.
Finally, each edge is associated with a few random variables $V_{\langle e_i \rangle}^{I_j}$, where each random variable describes the distribution on the edge during an interval and has a set of qualified trajectories $\mathbb{T}_{\langle e_i \rangle}^{I_j}$.

\subsubsection{Representing Univariate Distributions}
\label{sssec:repuni}

We represent distributions by histograms because they are able to approximate arbitrary distributions. In particular, a one-dimensional histogram is employed to represent a univariate distribution.
When $|\mathbb{T}_{\langle e_i \rangle}^{I_j}|>\beta$, we are able to obtain a multiset of cost values of the form $\langle\mathit{cost}, \mathit{perc}\rangle$, meaning that $\mathit{perc}$ percentage of the trajectories in $\mathbb{T}_{\langle e_i \rangle}^{I_j}$ took cost $\mathit{cost}$. We call this a \emph{raw cost distribution}.
A histogram then approximates the raw cost distribution as a set of pairs: $\{ \langle \mathit{bu}_i, \mathit{pr}_i \rangle\}$.
A bucket $\mathit{bu}_i=[l, u)$ is a range of travel costs, and $\mathit{pr}_i$ is the probability that the travel cost is in the range, and it holds that $\sum_{\mathit{pr}_i} \mathit{pr}_i=1$.

Given the number of buckets $b$, existing techniques, e.g., V-Optimal~\cite{DBLP:conf/vldb/JagadishKMPSS98}, are able to optimally derive a histogram based on a raw cost distribution such that the sum of errors between the derived histogram and the raw cost distribution is minimized.
However, selecting a global value for $b$ is non-trivial because the traffic on different edges, and even the traffic on the same edge during different intervals, may differ significantly. A self-tuning method is desired such that more buckets are used for edges or intervals with more complex traffic conditions.

To this end, we propose a simple yet effective approach to automatically identify the number of buckets based on the cost values $\mathit{costs}$ observed in the qualified trajectories $\mathbb{T}_{\langle e_i \rangle}^{I_j}$.
The procedure starts with $b=1$, i.e., using only one bucket, and computes an error value $E_b$. Next, it incrementally increases $b$ by 1 and computes a new error value.
Obviously, as the number of buckets increases, the error value keeps decreasing.
However, the error values often initially drop quickly, but then subsequently drop only slowly.
Based on this, the process stops when the error value of using $b$ does not lead to a significant decrease compared to the error value of using $b-1$. Then, $b-1$ is chosen as the bucket number. This yields a compact and accurate representation of the raw cost distribution.

The error value $E_b$ of using $b$ buckets is computed using $f$-fold cross validation~\cite{smyth2000model}. %
First, cost values $\mathit{costs}$ are randomly split into $f$ equal-sized partitions. Each time, we reserve the cost values in one partition, say the $k$-th partition, and use the cost values in the remaining $f-1$ partitions to generate a histogram with $b$ buckets using V-Optimal, denoted as $H_b^k=\{\langle \mathit{bu}_i$, $\mathit{pr}_i\rangle\}$. Next, we compute the raw cost distribution of the cost values in the reserved partition, denoted as $D^k=\{\langle \mathit{cost}_i$, $\mathit{perc}_i\rangle\}$. After that, we compute the squared error between $H_b^k$ and $D^k$: $SE(H_b^k, D^k)=\sum_{c\in\mathit{costs}} (H_b^k[c], D^k[c])^2$,
where $H_b^k[c]$ and $D^k[c]$ denote the probability of cost $c$ in the histogram $H_b^k$ and raw cost distribution $D^k$, respectively.
We repeat the procedure $f$ times---once for each partition. The error value of using $b$ buckets, i.e., $E_b$, is the average of the $f$ squared errors.
The pseudo code of the procedure is described in Algorithm~\ref{alg:auto} in Appendix~\ref{ap:code}.

\begin{figure}[h]
  \centering
  \subfigure[$E_b$ vs. $b$]{
    \label{fig:subfig:fig6xdata} 
    \includegraphics[width=0.47\columnwidth]{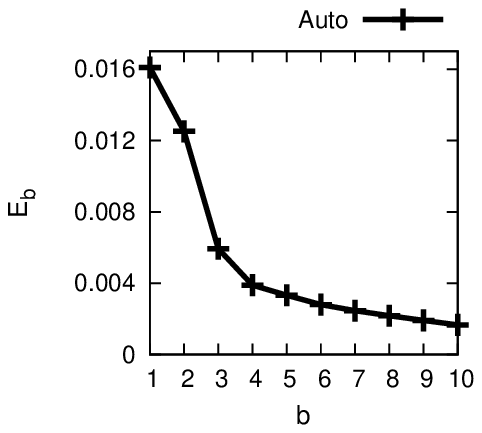}}
  \subfigure[Raw Dist. vs. Histogram]{
    \label{fig:subfig:fig6ydata} 
    \includegraphics[width=0.47\columnwidth]{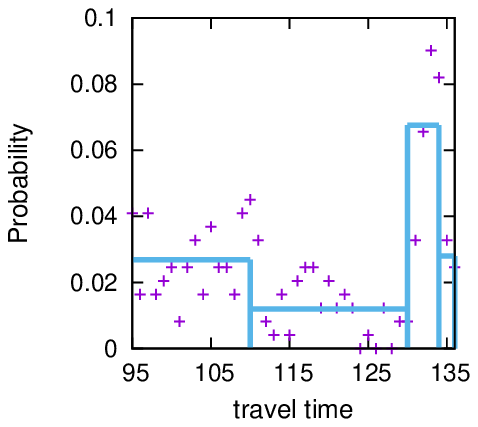}
    }
  \caption{Identifying the Number of Buckets}
  \label{fig:eb-vs-b} 
\end{figure}

Take the data in Figure~\ref{fig:introexample}(a) as an example. Figure~\ref{fig:eb-vs-b}(a) shows how the error $E_b$ decreases as the number of buckets $b$ increases.
First, $E_b$ decreases sharply and then slowly (i.e., when $b>4$).
Figure~\ref{fig:eb-vs-b}(b) shows the histogram using $b=4$ buckets and the original raw cost distribution.

\subsection{Deriving Joint Distributions for Paths}

Based on the distributions of unit paths, we employ a bottom-up procedure to derive joint distributions of paths with cardinalities more than one. In particular, the joint distributions of paths with cardinalities $k$, $k\geqslant 2$, are computed based on the joint distributions of paths with cardinalities $k-1$.

We first consider paths with cardinalities two. Given two unit paths $\mathcal{P}_a=\langle e_a \rangle$ and $\mathcal{P}_b=\langle e_b \rangle$, if their concatenation $\mathcal{P}=\mathcal{P}_a\circ\mathcal{P}_b$ is not empty, i.e., $\mathcal{P}=\langle e_a, e_b \rangle$, we check if a time interval $I_j$ exists during which more than $\beta$ trajectories occurred on path $\mathcal{P}$. If so, a random variable $V_{\mathcal{P} }^{I_j}=p(c_{e_a}, c_{e_b})$ is instantiated based on the trajectories $\mathbb{T}_{\mathcal{P}}^{I_j}$.

\begin{figure}[!hpt]
  \centering
  \subfigure[Raw Distribution]{
    \label{fig:subfig:actual} 
    \includegraphics[width=0.47\columnwidth]{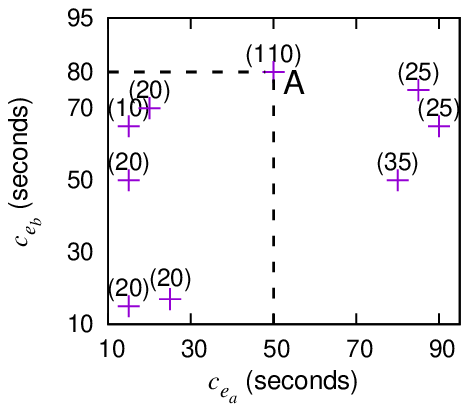}}
  \subfigure[2D Histogram]{
    \label{fig:subfig:appro} 
    \includegraphics[width=0.47\columnwidth]{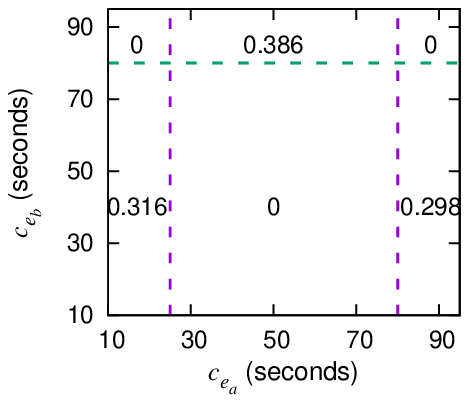}}
  \caption{An Example of Multiple Dimensional Histogram}
  \label{fig:histo-appr} 
\end{figure}

The travel cost distribution on a path $p(c_{e_a}, c_{e_b})$ is a joint distribution of the two variables $c_{e_a}, c_{e_b}$.
For example, Figure~\ref{fig:subfig:actual} shows a raw joint distribution. Point A indicates that 110 trajectories passed $e_a$ with cost 50~s and then $e_b$ with cost 80~s.

We use multi-dimensional histograms to describe joint distributions.
A multi-dimensional histogram is a set of hyper-bucket and probability pairs: $\{ \langle \mathit{hb}_i, \mathit{pr}_i \rangle\}$. %
A hyper-bucket $\mathit{hb}_i=\langle bu_i^1, \ldots bu_i^n\rangle $ consists of $n$ buckets that each corresponds to one dimension. Value $\mathit{pr}_i$ equals the probability that the travel costs on multiple edges are in the hyper-bucket, and it holds that $\sum_{\mathit{pr}_i} \mathit{pr}_i=1$.

To derive a multi-dimensional histogram, we automatically identify the optimal number of buckets for each dimension using the method from Section~\ref{sssec:repuni}. Next, we employ V-optimal to identify the optimal bucket boundaries on each dimension and thus obtain a set of hyper-buckets. Finally, we compute the probability for each hyper-bucket.
For example, Figure~\ref{fig:subfig:appro} shows a 2-dimensional histogram that corresponds to the joint distribution shown in Figure~\ref{fig:subfig:actual}. One dimension is partitioned into 3 buckets and the other one is partitioned into 2 buckets, yielding 6 hyper-buckets in the 2-dimensional histogram.

Next, we consider paths with cardinalities $k\geqslant 3$. Given two paths $\mathcal{P}_i$ and $\mathcal{P}_j$ with cardinalities $k-1$, if they share $k-2$ edges and can be combined into a path $\mathcal{P}=\langle e_1, e_2, \ldots, e_k \rangle $ with cardinality $k$, we check if a time interval exists during which more than $\beta$ trajectories occurred on path $\mathcal{P}$. If so, a random variable $V_{\mathcal{P}}^{I_j}=p(c_{e_1}, \ldots, c_{e_k})$ is instantiated based on the trajectories $\mathbb{T}_{\mathcal{P}}^{I_j}$.
The travel cost distribution $p(c_{e_1}, \ldots, c_{e_k})$ is a joint distribution on the $k$ variables $c_{e_1}, \ldots, c_{e_k}$.
V-optimal is also able to compute a $k$-dimensional histogram to represent the joint distribution $p(c_{e_1}, \ldots, c_{e_k})$.
This procedure continues until longer paths cannot be formed.

\subsection{Learned Random Variables}
To summarize, we let $\mathbf{C}_{\mathcal{P}}$ be a vector of variables $\langle c_{e_1}, c_{e_2}, \ldots, \\c_{e_k} \rangle$ that correspond to path $\mathcal{P}=\langle e_1, e_2, \ldots, e_k \rangle$.
By analyzing a large trajectory set, we are able to obtain a set of \emph{learned random variables} $\mathit{LRV}=\{V_{\mathcal{P}}^{I_j}\}$, where each random variable $V_{\mathcal{P}}^{I_j}$ describes the joint distribution of traversing path $\mathcal{P}$ during $I_j$. Specifically, \\
(i) if $\mathcal{P}=\langle e_i \rangle$ is a unit path, $V_{\mathcal{P}}^{I_j}$ is the distribution $p(\mathbf{C}_{\mathcal{P}})=p(c_{e_i})$, which is represented as a one-dimensional histogram;\\
(ii) if $\mathcal{P}=\langle e_1, e_2, \ldots, e_k \rangle$ is not a unit path, $V_{\mathcal{P}}^{I_j}$ is the joint distribution $p(\mathbf{C}_{\mathcal{P}})=p(c_{e_1}, c_{e_2}, \ldots, c_{e_k})$, represented as a multi-dimensional histogram.

We define the \emph{rank} of a learned random variable $V_{\mathcal{P}_j}^{I_j}$ as the cardinality of its path $|\mathcal{P}_j|$.
In the conventional paradigm, only random variables with rank one are considered, and they cannot capture the distribution dependencies among edges, which may be represented in trajectories.
In contrast, in the new paradigm, the random variables with rank greater than one fully capture the distribution dependencies among edges in a path with sufficiently many qualified trajectories.

\section{Travel Cost Distribution Estimation}
\label{sec:TCDE}

We perform travel cost estimation \emph{on-line}, in two steps. First, the joint distribution of path $\mathcal{P}$, which models the travel cost dependency among edges in $\mathcal{P}$, is computed. Second, the marginal distribution of path $\mathcal{P}$, which captures the cost distribution of traversing path $\mathcal{P}$, is derived based on the joint distribution of $\mathcal{P}$.

\subsection{The Joint Distribution of a Path}

The joint distribution of path $\mathcal{P}=\langle e_1, e_2, \ldots, e_n\rangle$ at $t$ is denoted $p(\mathbf{C}_{\mathcal{P}})=p(c_{ e_1}, c_{ e_2}, \ldots, c_{ e_n})$, where $c_{ e_i}$ ($1\leqslant i \leqslant n$) is a random variable representing the travel cost distribution of path $\langle e_i\rangle$.
We proceed to propose a method that is able to derive an accurate, estimated joint distribution $\hat{p}(c_{ e_1}, {c}_{ e_2},\cdots c_{ e_n})$ based on the learned random variable set $\mathit{LRV}$.
While we may be able to obtain multiple estimations of joint distributions using different combinations of learned random variables, we are only interested in the most accurate one.

\subsubsection{Candidate Random Variable Sets}
\label{sssec:crvs}

Following the principles of decomposable models~\cite{malvestuto1991approximating,bishop2007discrete,darroch1983additive}, we define a candidate random variable set based on the set of learned random variables $\mathit{LRV}$.
Formally, a candidate random variable set is a sequence of learned random variables, denoted as $\mathit{CRV}=\langle V_{\mathcal{P}_1}^{I_1}, V_{\mathcal{P}_2}^{I_2}, \ldots, V_{\mathcal{P}_m}^{I_m} \rangle$, that satisfies a \emph{spatial condition} and a \emph{temporal condition}.

\textbf{Spatial condition:} The condition is that the corresponding paths $\mathcal{P}_1$, $\mathcal{P}_2$, $\ldots$, $\mathcal{P}_m$ of the random variables in $\mathit{CRV}$ satisfy the following: \\
(i) $\mathcal{P}_1 \cup \mathcal{P}_2 \cup \ldots \cup \mathcal{P}_{m}$ is path $\mathcal{P}$; \\
and, in addition, one of the following two must be satisfied.\\
(ii) $(\mathcal{P}_1 \cup \mathcal{P}_2 \cup \ldots \cup \mathcal{P}_{h-1}) \cap (\mathcal{P}_h) = \emptyset$, where $1 < h\leqslant m$; \\
(iii) if $(\mathcal{P}_1 \cup \mathcal{P}_2 \cup \ldots \cup \mathcal{P}_{h-1}) \cap \mathcal{P}_h \neq \emptyset$ then $(\mathcal{P}_1 \cup \mathcal{P}_2 \cup \ldots \cup \mathcal{P}_{h-1}) \cap \mathcal{P}_h = \mathcal{P}_{h-1} \cap \mathcal{P}_h$, where $1 < h\leqslant m$.

Condition (i) ensures that the random variables' paths together cover path $\mathcal{P}$.
Condition (ii) implies that the two adjacent random variables' paths do not share any edges, meaning that the random variables are independent of each other.
Condition (iii) implies that only two adjacent random variables' paths may share edges. In this case, the edges in the paths are \emph{conditionally independent} given the shared edges.  $\Box$

\textbf{Temporal condition: } This condition requires that the corresponding intervals $I_1$, $I_2$, $\ldots$, $I_m$ of the random variables in $\mathit{CRV}$ are temporally close to the departure time $t$.

We distinguish two cases: the case for the first path $\mathcal{P}_1$ and the case for the remaining paths.
Since $\mathcal{P}_1$ must start with edge $e_1$, interval $I_1$ is considered as temporally close to the departure time if $I_1$ covers the departure time, i.e., $t\in I_1$.
Next, when $2\leqslant j \leqslant m$, checking whether interval $I_j$ is temporally close becomes complicated as the departure time from $\mathcal{P}_j$ is no longer the original departure time $t$---one has to spend some time to reach $\mathcal{P}_j$. To this end, we introduce a ``shift-and-enlarge'' procedure to progressively update the departure time for $\mathcal{P}_j$ and check whether interval $I_j$ is close to the updated departure time. The details of the shift-and-enlarge procedure are covered in Appendix~\ref{ap:shiftandenlarge}. $\Box$

To ease the following discussions, we introduce a running example. Consider $\mathcal{P}= \langle e_1, e_2, e_3, e_4, e_5\rangle$ at departure time $t$. Assume the learned random variable set is $\mathit{LRV}=\{V_{\langle e_1\rangle}^{I_1}$, $V_{\langle e_1, e_2\rangle}^{I_2}$, $V_{\langle e_1, e_2, e_3\rangle}^{I_3}$, $V_{\langle e_1, e_2, e_3, e_4\rangle}^{I_4}$,
$V_{\langle e_2\rangle}^{I_5}$, $V_{\langle e_2, e_3\rangle}^{I_6}$, $V_{\langle e_2, e_3, e_4\rangle}^{I_7}$,
$V_{\langle e_3\rangle}^{I_8}$, $V_{\langle e_3, e_4\rangle}^{I_9}$,
$V_{\langle e_4\rangle}^{I_{10}}$, $V_{\langle e_4, e_5\rangle}^{I_{11}}$,
$V_{\langle e_5\rangle}^{I_{12}}\}$.
And assume all the intervals are temporally close to the updated departure time.

Based on $\mathit{LRV}$, multiple candidate random variable sets can be constructed while satisfying both the spatial and temporal conditions, including the following two sets: $\mathit{CRV}_1=\langle V_{\langle e_1, e_2, e_3\rangle}^{I_3}$, $V_{\langle e_2, e_3, e_4\rangle}^{I_7}$, $V_{\langle e_5\rangle}^{I_{12}} \rangle$ and  $\mathit{CRV}_2=\langle V_{\langle e_1, e_2, e_3\rangle}^{I_3}$, $V_{\langle e_3, e_4\rangle}^{I_9}$, $V_{\langle e_5\rangle}^{I_{12}} \rangle$.

For candidate random variable set $\mathit{CRV}_1$, the first two random variables $V_{\langle e_1, e_2, e_3\rangle}^{I_3}$ and $V_{\langle e_2, e_3, e_4\rangle}^{I_7}$ satisfy spatial condition (iii), and the travel cost distributions on edges $e_1$ and $e_4$ are conditionally independent given edges $e_2$ and $e_3$. Further, random variables $V_{\langle e_2, e_3, e_4\rangle}^{I_7}$ and $V_{\langle e_5\rangle}^{I_{12}}$ are independent, since $\langle e_2, e_3, e_4\rangle$ and $\langle e_5\rangle$ are disjoint.

Based on the candidate random variable set $\mathit{CRV}_1$, the joint distribution of path $\mathcal{P}$ can be estimated as $\hat{p}_{\mathit{CRV}_1}(\mathbf{C}_{\mathcal{P}})=\hat{p}_{\mathit{CRV}_1}(c_{e_1}, c_{e_2}, c_{ e_3 }, c_{ e_4 }, c_{ e_5 })$.
Since $\langle e_5\rangle$ is independent of the remaining edges, we have
$
\hat{p}_{\mathit{CRV}_1}(\mathbf{C}_{\mathcal{P}})=p(c_{e_1}, c_{e_2}, c_{ e_3 }, c_{ e_4 })\cdot p(c_{ e_5 }).
$
Next, since the random variables on $e_1$ and $e_4$ are conditionally independent given $e_2$ and $e_3$, we have
$
\hat{p}_{\mathit{CRV}_1}(\mathbf{C}_{\mathcal{P}})=\frac{p(c_{e_1}, c_{ e_2 }, c_{ e_3 }) \cdot p(c_{ e_2 }, c_{ e_3 }, c_{ e_4 })}{p(c_{ e_2 }, c_{ e_3 }) }\cdot p(c_{e_5}).
$

Similarly, for candidate random variable set $\mathit{CRV}_2$, we are able to derive
$
\hat{p}_{\mathit{CRV}_2}(\mathbf{C}_{\mathcal{P}})=\frac{p(c_{e_1}, c_{ e_2 }, c_{ e_3 }) \cdot p( c_{ e_3 }, c_{ e_4 })}{p(c_{ e_3 }) }\cdot p(c_{e_5}).
$

Formally, a joint distribution of path $\mathcal{P}$ can be estimated based on each candidate random variable set $\mathit{CRV}$, as defined in Equation~\ref{eq:est}.
\begin{equation}
\label{eq:est}
 \begin{aligned}
 & \hat{p}_{\mathit{CRV}}(\mathbf{C}_{\mathcal{P}})
 & = & \frac{\prod_{V_{\mathcal{P}_i}^{I_i}\in \mathit{CRV}}
 V_{\mathcal{P}_i}^{I_i}}{\prod_{V_{\mathcal{P}_i}^{I_i}\in \mathit{CRV}} V_{\mathcal{P}_i \cap \mathcal{P}_{i-1}}^{I_i}} = \frac{\prod_{\mathcal{P}_i \in P_X}
 p(\mathbf{C}_{\mathcal{P}_i})}{\prod_{\mathcal{P}_j \in P_Y} p(\mathbf{C}_{\mathcal{P}_j}) }, \\
 \end{aligned}
\end{equation}
where $P_X=\cup_{V_{\mathcal{P}_i}^{I_i}\in \mathit{CRV}} \mathcal{P}_i$ and $P_Y=\cup_{V_{\mathcal{P}_i}^{I_i}\in \mathit{CRV}} \mathcal{P}_i \cap \mathcal{P}_{i-1}$.

\subsubsection{Optimal Candidate Random Variable Set}

Assume that we have a set $\mathit{SS}$ of all possible candidate random variable sets that satisfy the spatial and temporal conditions.
For each candidate random variable set $\mathit{CRV}\in \mathit{SS}$, we can compute an estimated joint distribution $\hat{p}_{\mathit{CRV}}$ using Equation~\ref{eq:est}. Next, we aim at identifying the optimal candidate random variable set $\mathit{CRV}_{opt}\in \mathit{SS}$ such that the derived estimated joint distribution $\hat{p}_{\mathit{CRV}_{opt}}(\mathbf{C}_{\mathcal{P}})$ is the most accurate estimation.

Assume that the true, but unknown, joint distribution is ${p}(\mathbf{C}_{\mathcal{P}})$. To measure how accurate an estimated distribution is from the true distribution, we employ Kullback-Leibler divergence of the estimated joint distribution and the true joint distribution, denoted as
$\mathit{KL}({p}(\mathbf{C}_{\mathcal{P}}), \hat{p}_{\mathit{CRV}}(\mathbf{C}_{\mathcal{P}}))$.
The smaller the divergence, the more accurate the estimated distribution is. Thus, identifying the optimal candidate random variable set can be achieved by solving the following optimization problem.

\begin{equation*}
 \begin{aligned}
 & \underset{ \mathit{CRV} \in \mathit{SS}}{\text{argmin}}
 & & \mathit{KL}({p}(\mathbf{C}_{\mathcal{P}}), \hat{p}_{\mathit{CRV}}(\mathbf{C}_{\mathcal{P}})) \\
 \end{aligned}
\end{equation*}

We provide three theorems that facilitate solving this optimization problem.
\begin{theorem}
\label{th:0}
If path $\mathcal{P}^\prime$ is a sub-path of path $\mathcal{P}$, we have
$
\sum_{\mathbf{C}_{\mathcal{P}}} p(\mathbf{C}_{\mathcal{P}})\log p(\mathbf{C}_{\mathcal{P}^\prime})=-H(\mathbf{C}_{\mathcal{P}^\prime}).
$
\end{theorem}

\begin{proof}
Since path $\mathcal{P}^\prime$ is a sub-path of path $\mathcal{P}$, path $\mathcal{P}$ can be represented as $\mathcal{P}=\mathcal{P}_{s}\circ \mathcal{P}^\prime \circ \mathcal{P}_{e}$, where $\mathcal{P}_{s}$ and $\mathcal{P}_{e}$ are the paths before and after path $\mathcal{P}^\prime$, respectively.
\begin{equation*}
 \begin{aligned}
  \lefteqn{\sum_{\mathbf{C}_{\mathcal{P}}} p(\mathbf{C}_{\mathcal{P}})\log p(\mathbf{C}_{\mathcal{P}^\prime})}\\
 &= & & \sum_{\mathbf{C}_{\mathcal{P}_s}, \mathbf{C}_{\mathcal{P}^\prime}, \mathbf{C}_{\mathcal{P}_e}} p(\mathbf{C}_{\mathcal{P}_s}, \mathbf{C}_{\mathcal{P}^\prime}, \mathbf{C}_{\mathcal{P}_e})\log p(\mathbf{C}_{\mathcal{P}^\prime})\\
 &= & & \sum_{\mathbf{C}_{\mathcal{P}^\prime}} \big(\log p(\mathbf{C}_{\mathcal{P}^\prime}) \sum_{\mathbf{C}_{\mathcal{P}_s}, \mathbf{C}_{\mathcal{P}_e}}p(\mathbf{C}_{\mathcal{P}_s}, \mathbf{C}_{\mathcal{P}^\prime}, \mathbf{C}_{\mathcal{P}_e}) \big)\\
 &= & & \sum_{\mathbf{C}_{\mathcal{P}^\prime}} p(\mathbf{C}_{\mathcal{P}^\prime})  \log p(\mathbf{C}_{\mathcal{P}^\prime}) = -H(\mathbf{C}_{\mathcal{P}^\prime}) \\
   \end{aligned}
 \end{equation*}

This completes the proof.
\end{proof}

\begin{theorem}
\label{th:1}
Given an estimated joint distribution $\hat{p}_{\mathit{CRV}}(\mathbf{C}_{\mathcal{P}})$, we have $\mathit{KL}(p(\mathbf{C}_{\mathcal{P}}), \hat{p}_{\mathit{CRV}}(\mathbf{C}_{\mathcal{P}}))=H_{\mathit{CRV}}(\mathbf{C}_{\mathcal{P}})-H(\mathbf{C}_{\mathcal{P}})$, where $H(\mathbf{C}_{\mathcal{P}})$ and $H_{\mathit{CRV}}(\mathbf{C}_{\mathcal{P}})$ are the entropies of random variable $\mathbf{C}_{\mathcal{P}}$ under distributions $p(\mathbf{C}_{\mathcal{P}})$ and $\hat{p}_{\mathit{CRV}}(\mathbf{C}_{\mathcal{P}})$, respectively.
\end{theorem}
\begin{proof}
 \begin{equation*}
 \begin{aligned}
  \lefteqn{\mathit{KL}(p(\mathbf{C}_{\mathcal{P}}), \hat{p}_{\mathit{CRV}}(\mathbf{C}_{\mathcal{P}}))}\\
&= & & \sum_{\mathbf{C}_{\mathcal{P}}}p(\mathbf{C}_{\mathcal{P}})\log\big(\frac{p(\mathbf{C}_{\mathcal{P}})}{\hat{p}_{\mathit{CRV}}(\mathbf{C}_{\mathcal{P}})}\big) \\
 &= & &-H(\mathbf{C}_{\mathcal{P}})-\sum_{\mathbf{C}_{\mathcal{P}}}p(\mathbf{C}_{\mathcal{P}})\log\ \hat{p}_{\mathit{CRV}}(\mathbf{C}_{\mathcal{P}}) \\
 &=& & -H(\mathbf{C}_{\mathcal{P}})-\sum_{\mathbf{C}_{\mathcal{P}}}{p}(\mathbf{C}_{\mathcal{P}})\big(\sum_{\mathcal{P}_i\in P_X}\log\ p(\mathbf{C}_{\mathcal{P}_i})\\
 &
 & & -\sum_{\mathcal{P}_j\in P_Y}\log\ p(\mathbf{C}_{\mathcal{P}_j})\big)   \text{~~~~~(due to Eq.~\ref{eq:est})}\\
 &=& & -H(\mathbf{C}_{\mathcal{P}})-\sum_{\mathcal{P}_i\in P_X}\big(\sum_{\mathbf{C}_{\mathcal{P}}}{p}(\mathbf{C}_{\mathcal{P}})\log\ p(\mathbf{C}_{\mathcal{P}_i})\big)\\
 &
 & & +\sum_{\mathcal{P}_j\in P_Y} \big(\sum_{\mathbf{C}_{\mathcal{P}}}{p}(\mathbf{C}_{\mathcal{P}}) \log\ p(\mathbf{C}_{\mathcal{P}_j}) \big) \\
    \end{aligned}
\end{equation*}
 \begin{equation*}
 \begin{aligned}
 &=
 &&-H(\mathbf{C}_{\mathcal{P}})+\sum_{\mathcal{P}_i\in P_X}H(\mathbf{C}_{\mathcal{P}_i})-\sum_{\mathcal{P}_j\in P_Y}H(\mathbf{C}_{\mathcal{P}_j})\\
 & & & \text{~~~~~~~~~~~~~~~~~~~~~~~~~~~~~~(due to Theorem~\ref{th:0}.)}\\
 &=
 &&H_{\mathit{CRV}}(\mathbf{C}_{\mathcal{P}})-H(\mathbf{C}_{\mathcal{P}})
 \end{aligned}
\end{equation*}

This completes the proof.
\end{proof}

Although the true joint distribution distribution $p(\mathbf{C}_{\mathcal{P}})$ is unknown, it is fixed. Thus, its entropy $H(\mathbf{C}_{\mathcal{P}})$ is also constant. It then follows from Theorem~\ref{th:1} that minimizing $\mathit{KL}(p(\mathbf{C}_{\mathcal{P}}), \hat{p}_{\mathit{CRV}}(\mathbf{C}_{\mathcal{P}}))$ is equivalent to minimizing $H_\mathit{CRV}(\mathbf{C}_{\mathcal{P}})$.
In other words, the estimated joint distribution of the optimal candidate random variable set, i.e., $\hat{p}_{\mathit{CRV}_{opt}}(\mathbf{C}_{\mathcal{P}})$, should have the least entropy among the estimated joint distributions of all candidate random variable sets in $\mathit{SS}$.

Before presenting the next theorem, we define a \emph{coarser} relationship between candidate random variable sets.
Suppose the random variables in $\mathit{CRV}_i$ concern paths in $P_i=\{\mathcal{P}_1, \mathcal{P}_2, \ldots, \mathcal{P}_m\}$ and those in $\mathit{CRV}_j$ concern paths in $P_j=\{\mathcal{P}_1^\prime, \mathcal{P}_2^\prime, \ldots, \mathcal{P}_n^\prime\}$.  Then, $\mathit{CRV}_i$ is coarser than $\mathit{CRV}_j$ if for each path $\mathcal{P}_a\in P_j$, there is a path in $\mathcal{P}_b\in P_i$ such that $\mathcal{P}_a$ is a sub-path of $\mathcal{P}_b$.

To illustrate, recall the example with $\mathit{CRV}_1$ and  $\mathit{CRV}_2$. Here, $\mathit{CRV}_1$ is coarser than $\mathit{CRV}_2$ because each path in $P_2=\{\langle e_1, e_2, e_3\rangle, \langle e_3, e_4\rangle, \langle e_5\rangle\}$ is a sub-path of a path in $P_1=\{\langle e_1, e_2, e_3\rangle, \langle e_2, e_3, e_4\rangle, \langle e_5\rangle\}$.

\begin{theorem}
\label{th:2}
Let $\mathit{CRV}$ and $\mathit{CRV}^\prime$ be candidate random variable sets. If $\mathit{CRV}$ is coarser than $\mathit{CRV}^\prime$, the entropy of $\hat{p}_{\mathit{CRV}}(\mathbf{C}_{\mathcal{P}})$ is smaller than that of $\hat{p}_{\mathit{CRV}^\prime}(\mathbf{C}_{\mathcal{P}})$, and thus $\mathit{CRV}$ is able to provide a more accurate joint distribution estimation than is $\mathit{CRV}^\prime$.
\end{theorem}

\begin{proof}
Since Theorem~\ref{th:1} holds for $\mathit{CRV}$ and $\mathit{CRV'}$, if
we are able to prove that entropy $H_{\mathit{CRV}}(\mathbf{C}_{\mathcal{P}})$ is smaller than entropy $H_{\mathit{CRV}^\prime}(\mathbf{C}_{\mathcal{P}})$, we have $\mathit{KL}(p(\mathbf{C}_{\mathcal{P}})$, $ \hat{p}_{\mathit{CRV}}(\mathbf{C}_{\mathcal{P}}))$ is smaller than $\mathit{KL}(p(\mathbf{C}_{\mathcal{P}})$, $ \hat{p}_{\mathit{CRV}^\prime}(\mathbf{C}_{\mathcal{P}}))$. Thus, $\mathit{CRV}$ is able to provide a more accurate joint distribution estimation than is $\mathit{CRV}^\prime$.

Now, we need to prove $\Delta=H_{\mathit{CRV}}(\mathbf{C}_{\mathcal{P}})-H_{\mathit{CRV}^\prime}(\mathbf{C}_{\mathcal{P}})$ is smaller than 0.

$\Delta=H_{\mathit{CRV}}(\mathbf{C}_{\mathcal{P}})-H_{\mathit{CRV}^\prime}(\mathbf{C}_{\mathcal{P}})$
\begin{equation*}
 \begin{aligned}
 &&\Delta=&\sum_{\mathcal{P}_i \in P_X}H(\mathbf{C}_{\mathcal{P}_i})-\sum_{\mathcal{P}_j \in P_Y}H(\mathbf{C}_{\mathcal{P}_j})\\
 &&&-\big( \sum_{\mathcal{P}_m \in P_{X}^\prime}H(\mathbf{C}_{\mathcal{P}_m})-\sum_{\mathcal{P}_n \in P_{Y}^\prime}H(\mathbf{C}_{\mathcal{P}_n})\big) \\
\end{aligned}
\end{equation*}

As $\mathit{CRV}$ is coarser than $\mathit{CRV}^\prime$, for each path $\mathcal{P}_m\in P_X^\prime$, there exists a path $\mathcal{P}_i\in P_X$ such that $\mathcal{P}_m$ is a sub-path of $\mathcal{P}_i$. Further, we are only interested in the cases where $|\mathcal{P}_m|<|\mathcal{P}_i|$; otherwise, the two paths are identical, and thus the difference between their entropies is zero.
For example, using the running example with $\mathit{CRV}_1$ and $\mathit{CRV}_2$, we are only interested in $\mathcal{P}_m=\langle e_3, e_4\rangle$ and $\mathcal{P}_i=\langle e_2, e_3, e_4\rangle$.

For each such $\mathcal{P}_m$ and $\mathcal{P}_i$ path pair, there exists a corresponding path pair $\mathcal{P}_n$ and $\mathcal{P}_j$ such that $\mathcal{P}_n$ is a sub-path of $\mathcal{P}_j$ and $|\mathcal{P}_n|<|\mathcal{P}_j|$.
Using the running example, we have $\mathcal{P}_n=\langle e_3\rangle$ and $\mathcal{P}_j=\langle e_2, e_3\rangle$.
We introduce notation $\mathcal{P}_{x-y}$ to denote the path $\mathcal{P}_{x}$ excluding edges in $\mathcal{P}_{y}$, and we have $\mathcal{P}_{i-m}=\mathcal{P}_{j-n}$. For example,  we have $\mathcal{P}_{i-m}=\langle e_2\rangle$ and $\mathcal{P}_{j-n}=\langle e_2\rangle$.

Based on the above, we have
\begin{equation*}
\begin{small}
    \begin{aligned}
    &\Delta&=\sum_{\mathcal{P}_m, \mathcal{P}_n, \mathcal{P}_i, \mathcal{P}_j}
          \big( H(\mathbf{C}_{\mathcal{P}_i}) - H(\mathbf{C}_{\mathcal{P}_m})
           -H(\mathbf{C}_{\mathcal{P}_j}) +H(\mathbf{C}_{\mathcal{P}_n})\big).\\
    \end{aligned}
\end{small}
\end{equation*}

According to the \emph{chain rule of conditional entropy}~\cite{cover2012elements}, we have
\begin{equation*}
 \begin{aligned}
& H(\mathbf{C}_{\mathcal{P}_i})=H(\mathbf{C}_{\mathcal{P}_m}) + H(\mathbf{C}_{\mathcal{P}_{i-m}}|\mathbf{C}_{\mathcal{P}_m}), \\
& H(\mathbf{C}_{\mathcal{P}_j})=H(\mathbf{C}_{\mathcal{P}_n}) + H(\mathbf{C}_{\mathcal{P}_{j-n}}|\mathbf{C}_{\mathcal{P}_n}). \\
\end{aligned}
\end{equation*}
As $\mathcal{P}_{i-m}=\mathcal{P}_{j-n}$, we have
\begin{equation*}
 \begin{aligned}
&\Delta&=\sum_{\mathcal{P}_m, \mathcal{P}_n, \mathcal{P}_i, \mathcal{P}_j}
       \big(H(\mathbf{C}_{\mathcal{P}_{i-m}}|\mathbf{C}_{\mathcal{P}_m})
       -H(\mathbf{C}_{\mathcal{P}_{i-m}}|\mathbf{C}_{\mathcal{P}_n})\big).\\
\end{aligned}
\end{equation*}

Based on the \emph{relationship between entropy and mutual information}~\cite{cover2012elements}, we have
\begin{equation*}
 \begin{aligned}
& H(\mathbf{C}_{\mathcal{P}_{i-m}}|\mathbf{C}_{\mathcal{P}_m})=H(\mathbf{C}_{\mathcal{P}_{i-m}}) - I(\mathbf{C}_{\mathcal{P}_{i-m}}; \mathbf{C}_{\mathcal{P}_m}), \\
& H(\mathbf{C}_{\mathcal{P}_{i-m}}|\mathbf{C}_{\mathcal{P}_n})=H(\mathbf{C}_{\mathcal{P}_{i-m}}) - I(\mathbf{C}_{\mathcal{P}_{i-m}}; \mathbf{C}_{\mathcal{P}_n}); \\
\end{aligned}
\end{equation*}
and thus we have
\begin{equation*}
 \begin{aligned}
&\Delta&=\sum_{\mathcal{P}_m, \mathcal{P}_n, \mathcal{P}_i, \mathcal{P}_j}
       \big(I(\mathbf{C}_{\mathcal{P}_{i-m}}; \mathbf{C}_{\mathcal{P}_n})
       -I(\mathbf{C}_{\mathcal{P}_{i-m}}; \mathbf{C}_{\mathcal{P}_m})\big).\\
\end{aligned}
\end{equation*}

Recall that $\mathcal{P}_n$ is a sub-path of $\mathcal{P}_m$ and $|\mathcal{P}_n|<|\mathcal{P}_m|$.
Based on the \emph{chain rule of mutual information}~\cite{cover2012elements}, we have
\begin{equation*}
 \begin{aligned}
I(\mathbf{C}_{\mathcal{P}_{i-m}}; \mathbf{C}_{\mathcal{P}_m})=I(\mathbf{C}_{\mathcal{P}_{i-m}}; \mathbf{C}_{\mathcal{P}_n}, \mathbf{C}_{\mathcal{P}_{m-n}})\\
=I(\mathbf{C}_{\mathcal{P}_{i-m}}; \mathbf{C}_{\mathcal{P}_n}) + I(\mathbf{C}_{\mathcal{P}_{i-m}};\mathbf{C}_{\mathcal{P}_{m-n}}|\mathbf{C}_{\mathcal{P}_n}).\\
\end{aligned}
\end{equation*}
Therefore, we have
\begin{equation*}
 \begin{aligned}
&\Delta&=\sum_{\mathcal{P}_m, \mathcal{P}_n, \mathcal{P}_i, \mathcal{P}_j}
       I(\mathbf{C}_{\mathcal{P}_{i-m}};\mathbf{C}_{\mathcal{P}_{m-n}}|\mathbf{C}_{\mathcal{P}_n}).\\
\end{aligned}
\end{equation*}
Since the mutual information on two different non-independent sets of random variables (i.e., $\mathbf{C}_{\mathcal{P}_{i-m}}$ and $\mathbf{C}_{\mathcal{P}_{m-n}}$) is positive, we have $\Delta<0$.

Therefore, the coarser $\mathit{CRV}$ is able to provide a more accurate estimated distribution than  is $\mathit{CRV}^\prime$.
\end{proof}

\subsubsection{Identifying $\mathit{CRV}_{opt}$}
\label{sssec:idenopt}

In continuation of Theorem~\ref{th:2}, we propose a method to identify the optimal candidate random variable set $\mathit{CRV}_{opt}$. The basic idea is to try to include the learned random variables with high ranks while maintaining the spatial and temporal conditions defined in Section~\ref{sssec:crvs} for being a candidate random variable set.

Given the $\mathit{LRV}$, we select a subset of random variables that are spatially and temporally relevant to the path $\mathcal{P}=\langle e_1, e_2, \ldots, e_n\rangle$ and time $t$. A random variable $V_{\mathcal{P}_j}^{I_j}$ is spatially relevant if its path $\mathcal{P}_j$ is a sub-path of $\mathcal{P}$. A random variable is temporally relevant if its interval $I_j$ is temporally close to the departure time $t$ according to the shift-and-enlarge procedure.

To simplify the following discussions, we omit the interval parts of the random variables and write $V_{\mathcal{P}_j}^{I_j}$ as $V_{\mathcal{P}_j}$.

Next, we organize the selected random variables into a two-dimensional ``candidate'' array. The $i$-th row contains the random variables whose paths starting with edge $e_i$. The $j$-th columns contains the rank $j$ random variables.
The array in Table~\ref{table:greedily-expansion} gives the random variables that are relevant for path $\mathcal{P}=\langle e_1, e_2, e_3, e_4, e_5\rangle$ in our running example.

\begin{table}[!htb]
\centering
\small
\begin{tabular}{ |c||c|c|c|c| }
\hline
& $\mathit{rank}=1$& $\mathit{rank}=2$ & $\mathit{rank}=3$ &$\mathit{rank}=4$\\\hline \hline
$e_1$&$V_{\langle e_1 \rangle}$& $V_{\langle e_1, e_2\rangle}$ & $V_{\langle e_1, e_2, e_3\rangle}$ & \cellcolor{blue!25}$V_{\langle e_1, e_2, e_3, e_4\rangle}$\\
$e_2$&$V_{\langle e_2\rangle}$&  $V_{\langle e_2, e_3\rangle}$ & $V_{\langle e_2, e_3, e_4\rangle}$ &$ $\\
$e_3$&$V_{\langle e_3\rangle}$&  $V_{\langle e_3, e_4\rangle}$ & $ $ &\\
$e_4$&$V_{\langle e_4\rangle}$&  \cellcolor{blue!25} $V_{\langle e_4, e_5\rangle}$ & &\\
$e_5$&$V_{\langle e_5\rangle }$&  & &\\
\hline
\end{tabular}
\caption{Example Candidate Array}
\label{table:greedily-expansion}
\end{table}

The method checks the array row by row. For each row, it constructs a candidate random variable set based on the random variable with the highest rank (i.e., the right-most element in the row), which is called the base variable.
For example, when checking the first row in Table~\ref{table:greedily-expansion}, we construct a candidate random variable set based on the base variable $V_{\langle e_1, e_2, e_3, e_4\rangle}$.
According to Theorem~\ref{th:2}, we can safely ignore all the random variables whose paths are sub-paths of $\langle e_1$, $e_2$, $e_3$, $e_4\rangle$. The remaining ones are $V_{\langle e_4, e_5\rangle}$ and $V_{\langle e_5\rangle }$. We consider $V_{\langle e_4, e_5\rangle}$ first since it has a longer path than does $V_{\langle e_5\rangle }$ and thus may produce a coarser candidate random variable set. Since $\{V_{\langle e_1, e_2, e_3, e_4\rangle}$, $V_{\langle e_4, e_5\rangle}\}$ satisfies the conditions stated in Section~\ref{sssec:crvs}, it is the candidate random variable set based on the first row.%

After this procedure, we get a set of candidate random variable sets, which we call $M$.
The optimal candidate random variable set $\mathit{CRV}_{opt}$ is the one with the least entropy.

\begin{theorem}
\label{th:3}
$\mathit{CRV}_{opt}=\arg\min_{\mathit{CRV} \in M}H_{\mathit{CRV}}(\mathbf{C}_{\mathcal{P}})$
\end{theorem}
\begin{proof}
We prove the theorem by contradiction. If Theorem~\ref{th:3} is incorrect, we cannot identify $\mathit{CRV}_{opt}$ by considering the candidate random variable sets in M. This happens only if $\mathit{CRV}_{opt}$ does not exist in $M$. Then, it must not contain any random variable with the highest rank for each row in the candidate array. Otherwise, it must be found by the presented procedure and thus belongs to $M$.

Let us assume that $\mathit{CRV}_{opt}$ contains a random variable $V_{\mathcal{P}}$ and a random variable $V_{\mathcal{P}^*}$ exists in the candidate array such that $\mathcal{P}$ is a sub-path of $\mathcal{P}^*$ and $|\mathcal{P}|<|\mathcal{P}^*|$. By replacing $V_{\mathcal{P}}$ by $V_{\mathcal{P}^*}$, the conditions given in Section~\ref{sssec:crvs} still hold, and thus we get a new candidate random variable set $\mathit{CRV}_{opt}^*$. Further, $\mathit{CRV}_{opt}^*$ is coarser than $\mathit{CRV}_{opt}$. According to Theorem~\ref{th:2}, $\mathit{CRV}_{opt}^*$ should have a smaller entropy. This contradicts the assumption that $\mathit{CRV}_{opt}$ has the least entropy.
\end{proof}

\subsection{The Marginal Distribution of a Path}
\label{ssec:marginaldist}

The identified optimal candidate random variable set $\mathit{CRV}_{opt}$ enables accurate estimation of the joint distribution of a path which fully captures the dependencies among edges in the path.
Recall that we are interested in knowing the marginal distribution of a path $p(V_{\mathcal{P}})$, where $V_{\mathcal{P}}$ is a univariate random variable indicating the travel cost of path $\mathcal{P}$.
We proceed to derive $p(V_{\mathcal{P}})$ based on the joint distribution of a path $\hat{p}_{\mathit{CRV}_{opt}}(\mathbf{C}_{\mathcal{P}})$ using Equation~\ref{eq:jointma}.
\begin{equation}
\label{eq:jointma}
 \begin{aligned}
 &p(V_{\mathcal{P}}=x)
 &=& \sum_{c_1+\ldots+c_n=x}\hat{p}_{\mathit{CRV}_{opt}}(c_{e_1}=c_1,\ldots,c_{e_n}=c_n). \\
 \end{aligned}
\end{equation}

The estimated joint distribution of a path $\hat{p}_{\mathit{CRV}_{opt}}(\mathbf{C}_{\mathcal{P}})$ is represented as a multi-dimensional histogram. Recall that a multi-dimensional histogram is of the form $\{\langle \mathit{hb}_i, \mathit{pr}_i \rangle \}$, where hyper-bucket $\mathit{hb}_i=\langle bu_i^1, \ldots bu_i^n\rangle $ consists of $n$ buckets, each corresponding to one dimension.

Take a simple example that concerns the joint distribution of a path with cardinality two $\mathcal{P}_1=\langle e_7, e_8 \rangle$, as shown in Table~\ref{table:fe1e2}, which has four hyper-buckets.
The upper, left hyper-bucket $\langle [20, 30), [20, 40)\rangle$ has value 0.3.
This means that when going through path $\langle e_7, e_8 \rangle$, the probability that the travel cost on edge $e_7$ is between 20~s and 30~s and the travel cost on edge $e_8$ is between 20~s and 40~s is 0.3.
\begin{table}[!htb]
\centering
\small
\begin{tabular}{ |c|c|c|c| }
\hline
  & $c_{e_7}\in [20,30)$ & $c_{e_7}\in[30,50)$ \\
\hline
$c_{e_8}\in[20,40)$ & 0.30 & 0.25\\
$c_{e_8}\in[40,60)$ & 0.20 & 0.25 \\
\hline
\end{tabular}
\caption{An Example $\hat{p}_{\mathit{CRV}_{opt}}(\mathbf{C}_{{\langle e_7, e_8 \rangle }})$}\label{table:fe1e2}
\end{table}

Next, we introduce two functions. Function $\mathit{HB2BU}: \mathit{HB} \rightarrow \mathit{BU}$ takes as input a hyper-bucket and outputs a bucket, where $\mathit{HB}$ and $\mathit{BU}$ indicate the sets of all possible hyper-buckets and buckets, respectively.
Specifically, given an hyper-bucket $\mathit{hb}_i=\langle bu_i^1, \ldots bu_i^n\rangle$, function $\mathit{HB2BU}$ outputs a bucket $bu=[\sum_{j=1}^{n} bu_i^{j}.l, \sum_{j=1}^{n} bu_i^{j}.u)$. The output bucket's upper (lower) bound is the sum of the upper (lower) bounds of the buckets in the hyper-bucket.
For example, after calling function $\mathit{HB2BU}$ on hyper-bucket $\langle [20, 30), [20, 40)\rangle$ in Table~\ref{table:fe1e2}, we obtain a bucket $[40, 70)$.

Function $\mathit{HB2BU}$ is called on each hyper-bucket $\mathit{hb}_i$ in the multi-dimensional histogram representing the joint distribution. Then, we get a set of (bucket, probability) pairs $\{\langle \mathit{HB2BU}(\mathit{hb}_i), \mathit{pr}_i \rangle \}$ representing the marginal distribution. For example, Table~\ref{table:afterHB2BU} shows the corresponding marginal distribution of the joint distribution that is shown in Table~\ref{table:fe1e2}.

\begin{table}[!htb]
\centering
\small
\begin{tabular}{ |c|c|c|c|c|}
\hline
$[40, 70)$ & $[60, 90)$ & $[50, 90)$ & $[70, 110)$\\
\hline
0.30 & 0.20 & 0.25 & 0.25 \\
\hline
\end{tabular}
\caption{$p(V_{\mathcal{P}_1})$, after calling $\mathit{HB2BU}$}\label{table:afterHB2BU}
\end{table}

The buckets in the obtained marginal distribution may overlap. For example, the first two buckets in Table~\ref{table:afterHB2BU}, $[40, 70)$ and $[60, 90)$, overlap. We define a function $\mathit{Rearrange}$ to rearrange the buckets such that buckets are disjoint, and each rearranged bucket is associated with an adjusted probability.
Formally, we have $\mathit{Rearrange}: \mathit{HP} \times \mathit{HP} \rightarrow 2^{\mathit{HP}}$, where $\mathit{HP}$ is a set of (bucket, probability) pairs. Given two such pairs $\langle bu_i, pr_i\rangle$ and $\langle bu_j, pr_j\rangle$, $\mathit{Rearrange}$ produces a set of (bucket, probability) pairs according to the following conditions.
If buckets $bu_i$ and $bu_j$ are disjoint, the set that consists of the two input pairs is returned.
If buckets $bu_i$ and $bu_j$ intersect, range $[\min(bu_i.l, bu_j.l), \max(bu_i.u, bu_j.u))$ is split into three buckets according to $bu_i.l$, $bu_j.l$, $bu_i.u$, and $bu_j.u$, and each bucket is assigned an adjusted probability.

Consider the first two (bucket, probability) pairs shown in Table~\ref{table:afterHB2BU}, i.e., $\langle [40, 70), 0.3 \rangle$ and $\langle [60, 90), 0.2 \rangle$. Since the two buckets intersect, range $[40, 90)$ is split into $[40, 60)$, $[60, 70)$, and $[70, 90)$.
In a histogram, the probability in each bucket is assumed to be uniformly distributed, so each bucket is assigned an adjusted probability as follows. Bucket $[40, 60)$ is given probability $\frac{|[40, 60)|}{|[40, 70)|} \cdot 0.3=0.2$, bucket $[60, 70)$ is given probability $\frac{|[60, 70)|}{|[40, 70)|} \cdot 0.3+$$\frac{|[60, 70)|}{|[60, 90)|} \cdot 0.2=0.17$, and bucket $[70, 90)$ is associated with probability $\frac{|[70, 90)|}{|[60, 90)|} \cdot 0.2=0.13$.
Calling function $\mathit{Rearrange}$ on these pairs repeatedly produces the final one-dimensional histogram that represents the corresponding marginal distribution. The final marginal distribution for the example is shown in Table~\ref{table:afterRe}.

\begin{table}[!htb]
\centering
\small
\begin{tabular}{ |c|c|c|c|c|c|}
\hline
$[40, 50)$ & $[50, 60)$ & $[60, 70)$ & $[70, 90)$ & $[90, 110)$ \\
\hline
0.1000 & 0.1625 & 0.2325 & 0.3800 & 0.1250\\
\hline
\end{tabular}
\caption{$p(V_{\mathcal{P}_1})$, after calling $\mathit{Rearrange}$}\label{table:afterRe}
\end{table}

The pseudo code of the whole procedure is shown in Algorithm~\ref{alg:joint-marginal} in Appendix~\ref{ap:code}.

\section{Empirical Study}
\label{sec:exp}

\subsection{Experimental Setup}

\textbf{Road networks:}
Two road networks are used in our experiments.
The Aalborg road network $N_{1}$ has 20,195 vertices and 41,276 edges, and
the average length of an edge is 
250 m.
The Beijing road network (within the 6-th ring road) $N_{2}$ 
has 28,342 vertices and 38,577 edges, and the average length of an edge is 200 m. 
Road network $N_{1}$ is obtained from OpenStreetMap and contains all roads, while road network $N_{2}$ is obtained from the Beijing traffic management bureau and contains only highways and main roads.

\textbf{Trajectories:} 
Two GPS record data sets are used. The first, $D_1$, contains $37$ million GPS records that occurred in Aalborg during January 2007 to December 2008. The sampling rate is 1~Hz (i.e., one record per second). The second, $D_2$, contains more than $50$ billion GPS records that occurred in Beijing from September 2012 to November 2012. The sampling rate is at least 0.2~Hz (i.e., at least one record per 5 seconds).
%
%
%
%
We apply a well-known method~\cite{DBLP:conf/gis/NewsonK09} to map match the GPS records.

\textbf{Travel Costs:} We consider two time-varying, uncertain travel costs---travel time and GHG emissions. The results on GHG emissions are included in Appendix~\ref{ap:ghg}.

\textbf{Parameters:} We vary parameters according to Table~\ref{table:defaulI_parameters}, where default values are shown in bold.
Specifically, we vary the finest time interval $\alpha$ (see Section~\ref{ssec:distunitpath}) from 15 to 120 minutes.
We vary the qualified trajectory count threshold $\beta$ from $15$ to $60$.
%
%
We vary the cardinality of a query path from 5 to 100. Parameter $f$, used in Section~\ref{sssec:repuni}, is set to 10, i.e., 10-fold cross validation.
%
%
\begin{table}[!htb]
\centering
\small
\begin{tabular}{ |c|c| }
\hline
Parameters & Values \\\hline
$\alpha$ (min) & 15, \textbf{30}, 45, 60 \\
$\beta$ & 15, \textbf{30}, 45, 60 \\
$|\mathcal{P}_{\mathit{query}}|$ & 5, 10, 15, \textbf{20}, 40, 60, 80, 100\\
\hline
\end{tabular}
\caption{Parameter Settings}\label{table:defaulI_parameters}
\end{table}

%

\textbf{Implementation Details:}
All algorithms are implemented in Python 2.7 under Linux Ubuntu 14.04.
All experiments are conducted on a modern server with 512 GB main memory and 64 2.3 GHz 8-core AMD Opteron(tm) 6376 CPUs.

\subsection{Experimental Results}
\subsubsection{Learned Random Variables}
\label{ssec:LRV}

We conduct a series of experiments to obtain insight into different aspects of the learned random variables. The random variables derived from speed limits are excluded from the following discussions.

\begin{figure*}[htbp]
\centering
\begin{tabular}{@{}c@{}c@{}c@{}c@{}}
    \includegraphics[width=0.48\columnwidth,height=3.2cm]{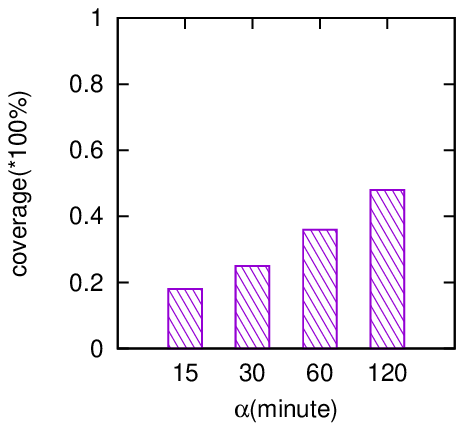}
    &
    \includegraphics[width=0.48\columnwidth,height=3.2cm]{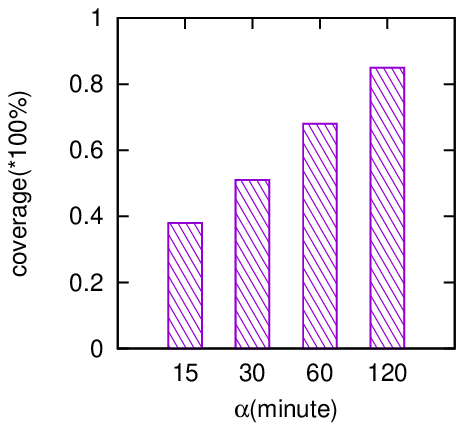}
    &
    \includegraphics[width=0.48\columnwidth,height=3.2cm]{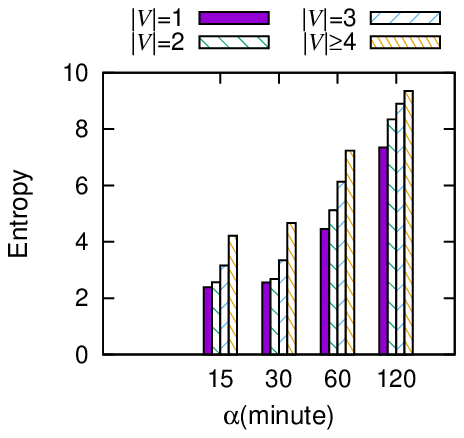}
    &
    \includegraphics[width=0.48\columnwidth,height=3.2cm]{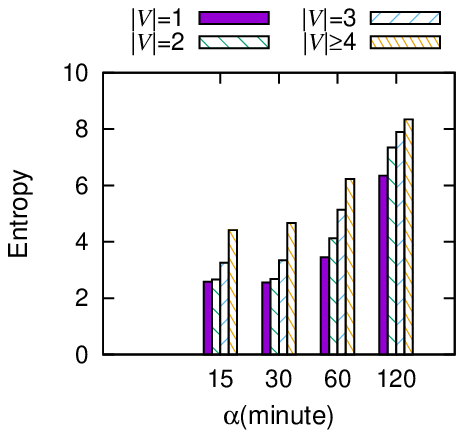}
    \\
    \small(a) Coverage, $D_1$
    &
    \small(b) Coverage, $D_2$
    &
    \small(c) Entropy, $D_1$
    &
    \small(d) Entropy, $D_2$
\end{tabular}
\caption{Effect of $\alpha$} \label{fig:alpha}
\end{figure*}

\textbf{Number of Learned Random Variables: } First, we vary parameter $\alpha$ from 15 to 120 minutes and study the effect of $\alpha$ on the number of learned random variables. A large $\alpha$ means that more trajectories become qualified trajectories. Thus, random variables with larger ranks can be learned.
We define an edge set $E^\prime=\cup_{V_{\mathcal{P}_i}^{I_i}} P_i$ that consists of all edges in paths of all the learned random variables. Coverage is defined as the ratio between $|E^\prime|$ and $|E^{\prime\prime}|$, where $E^{\prime\prime}$ is a set of edges that has at least one GPS record.
Figures~\ref{fig:alpha} (a) and (b) show that as $\alpha$ increases, the coverage increases as well on both data sets. 
However, the coverage ratio remains below $70$\% for $\alpha=120$. This is because the GPS data is skewed---some edges only have few GPS records during a day. %

Although a large $\alpha$ enables more learned random variables, the learned random variables may be inaccurate since traffic may change significantly during a long interval, e.g., two hours.
We report the average entropies of the learned random variables when varying $\alpha$ in Figures~\ref{fig:alpha} (c) and (d). According to Theorem~\ref{th:1}, variables with smaller entropy lead to more accurate joint distribution estimates. The figures 
show that using $\alpha=30$ does not significantly increase the entropy compared to using $\alpha=15$. Figures~\ref{fig:alpha} (a) and (b) show that $\alpha=30$ gives a clear increase in the number learned random variables compared to $\alpha=15$.
This suggests that $\alpha=30$ provides a good trade-off between the accuracy of the random variables and the numbers of random variables. Thus, we choose $\alpha=30$ as the default value in subsequent experiments.

\begin{figure*}[htbp]
   \begin{minipage}[b]{0.5\textwidth}
     \subfigure[$D_1$]{
     \includegraphics[width=0.49\textwidth,height=3.2cm]{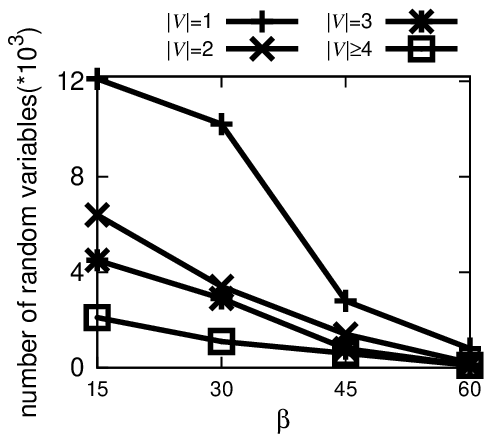}}
  \subfigure[$D_2$]{
    \includegraphics[width=0.49\textwidth,height=3.2cm]{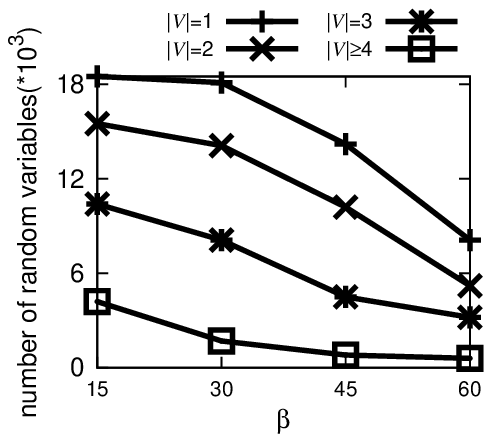}}
     \caption{Effect of $\beta$}
          \label{fig:betabeta}
   \end{minipage}
   \hfill
   \begin{minipage}[b]{0.5\textwidth}
       \centering
  \subfigure[$D_1$]{
    \includegraphics[width=0.48\textwidth,height=3.2cm]{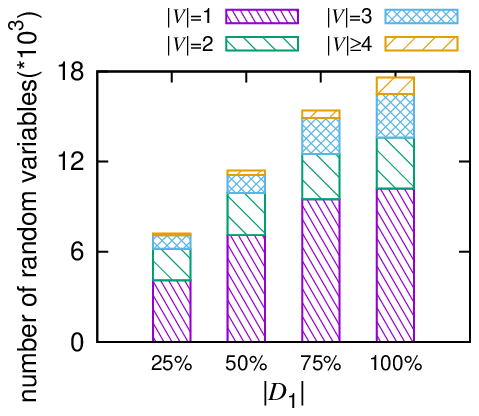}}
  \subfigure[$D_2$]{
     \includegraphics[width=0.48\textwidth,height=3.2cm]{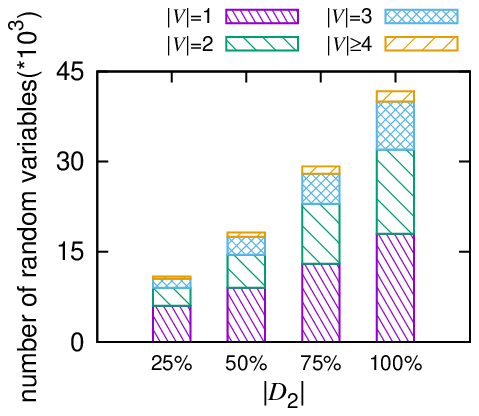}}
    \caption{Effect of the Count of Trajectories}
    \label{fig:rich-info}
   \end{minipage}
\end{figure*}

Next, we investigate the effect of parameter $\beta$. Intuitively, we prefer a large $\beta$ since having more trajectories enables accurate learned random variables.
However, as shown in Figures~\ref{fig:betabeta}(a) and (b), as $\beta$ increases, the number of learned random variables drops.
This occurs because a large $\beta$ requires more qualified trajectories to be available, which is less likely.
We choose $\beta=30$ as the default value because the number of learned random variables is only slightly less than that of $\beta=15$ while $\beta=30$ achieves more accurate random variables.

\begin{figure*}[htbp]
  \centering
  \subfigure[\textit{OCRV}]{
    \label{fig:overall-eva:odha}
    \includegraphics[width=0.23\textwidth,height=3.2cm]{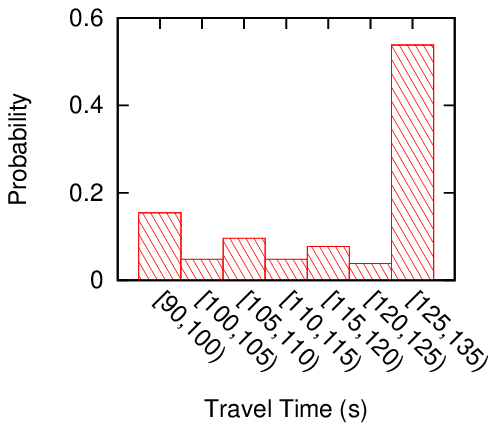}}
  \subfigure[\emph{LB}]{
        \label{fig:overall-eva:CV}
        \includegraphics[width=0.23\textwidth,height=3.2cm]{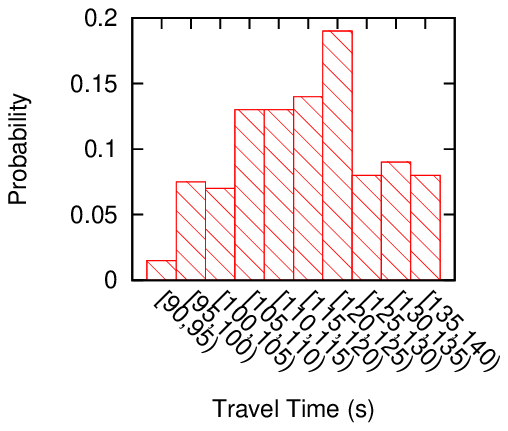}}
  \subfigure[\textit{HP}]{
    \label{fig:overall-eva:HP}
    \includegraphics[width=0.23\textwidth,height=3.2cm]{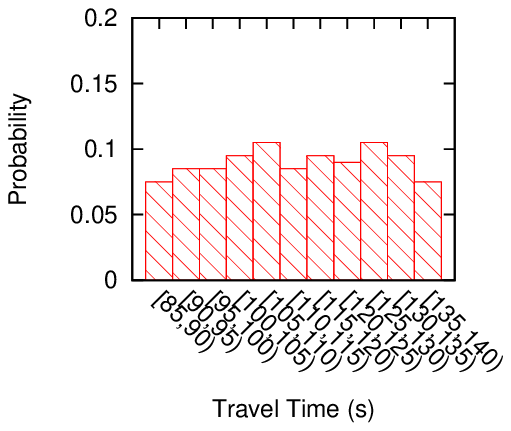}}
  \subfigure[\textit{CRV}]{
        \label{fig:overall-eva:DC}
        \includegraphics[width=0.23\textwidth,height=3.2cm]{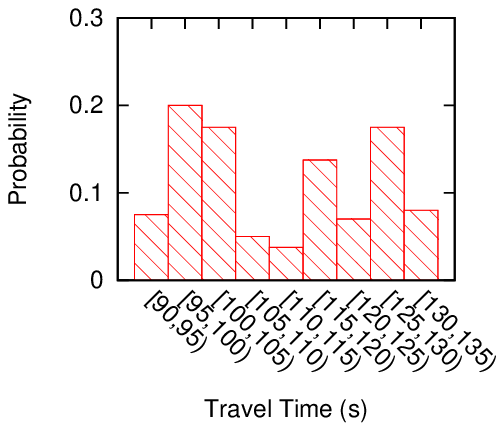}}
  \caption{Accuracy Comparison on a Particular Path}
  \label{fig:overall-detailed}
\end{figure*}

\begin{figure*}[htbp]
  \centering
  \subfigure[$D_1$]{
    \label{fig:aal-kl}
    \includegraphics[width=0.23\textwidth,height=3.2cm]{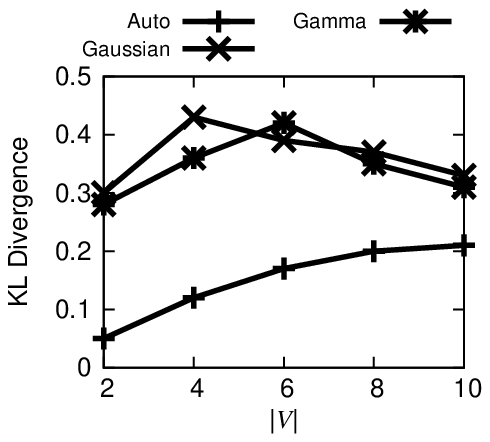}}
  \subfigure[$D_2$]{
    \label{fig:bj-kl}
    \includegraphics[width=0.23\textwidth,height=3.2cm]{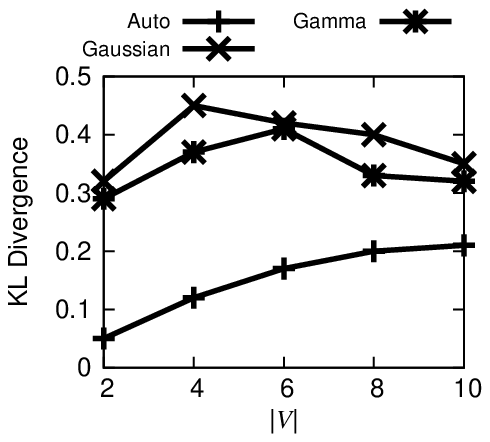}}
  \subfigure[Accuracy]{
    \label{fig:MV-opt:accur}
    \includegraphics[width=0.23\textwidth,height=3.2cm]{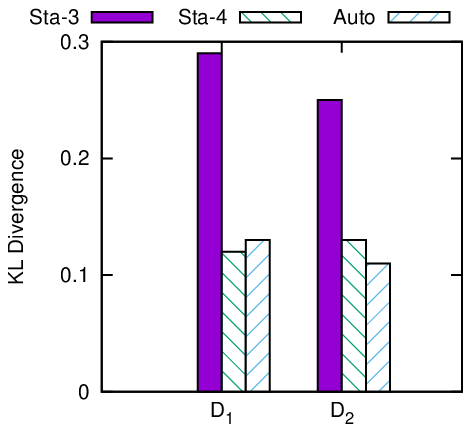}}
  \subfigure[Compression Ratio]{
    \label{fig:MV-opt:saved-space}
    \includegraphics[width=0.23\textwidth,height=3.2cm]{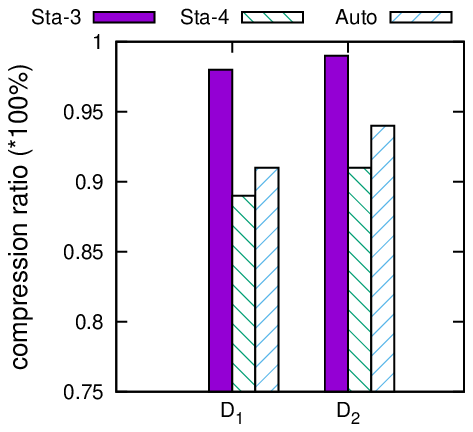}}
  \caption{Performance of Multi-Dimensional Histograms}
  \label{fig:MV-opt}
\end{figure*}

\begin{figure*}[htbp]
   \begin{minipage}[b]{0.5\textwidth}
     \subfigure[$D_1$]{
    \label{fig:overall-eva:d1}
    \includegraphics[width=0.48\textwidth]{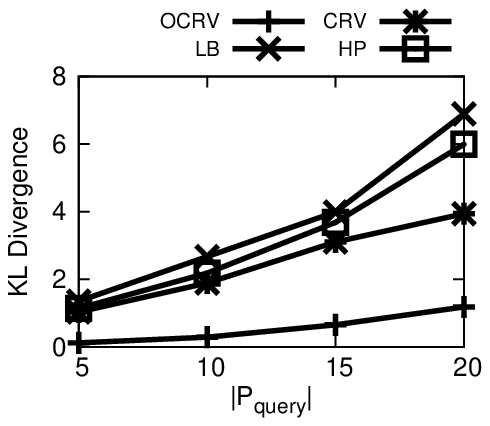}}
  \subfigure[$D_2$]{
    \label{fig:overall-eva:d2}
    \includegraphics[width=0.48\textwidth]{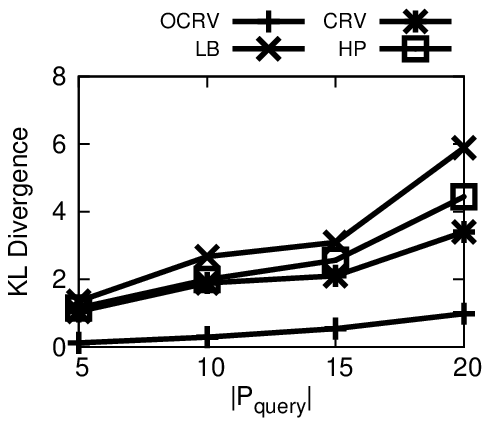}}
  \caption{Accuracy Comparison With The Ground Truth}
  \label{fig:overall-com}
   \end{minipage}
   \hfill
   \begin{minipage}[b]{0.5\textwidth}
       \centering
  \subfigure[$D_1$]{
    \label{fig:joint-prob:d1}
    \includegraphics[width=0.48\textwidth]{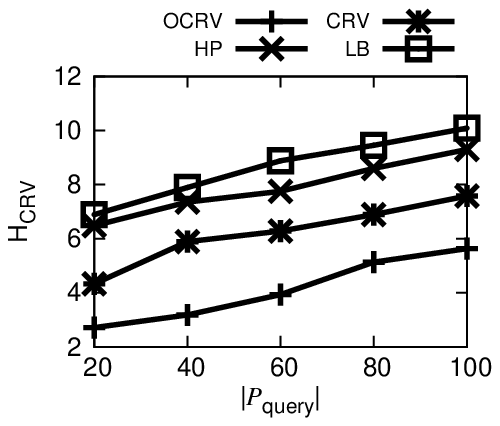}}
  \subfigure[$D_2$]{
    \label{fig:joint-prob:d2}
    \includegraphics[width=0.48\textwidth]{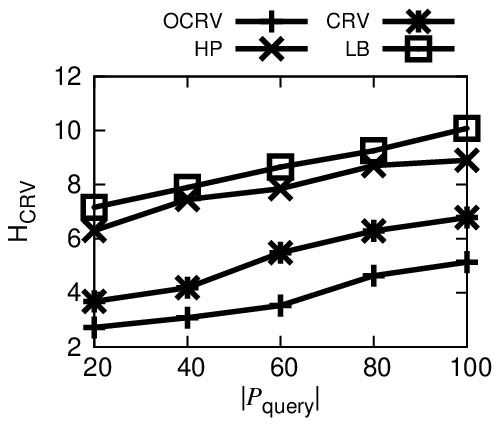}}
  \caption{Entropy Comparison}
  \label{fig:joint-prob}
   \end{minipage}
\end{figure*}

Additionally, we also study the number of the learned random variables w.r.t.\ the number of trajectories. To vary the sizes of trajectories, we use 25\%, 50\%, 75\%, and 100\% of the trajectories in $D_1$ and $D_2$, respectively.
Figures~\ref{fig:rich-info} (a) and (b) show that the number of learned random variables increases as the number of trajectories increases. We also see that the number of learned random variables with large ranks increases steadily.
This occurs because the more trajectories, the more likely it is to find long paths with more than threshold $\beta$ qualified trajectories, thus resulting in random variables with large ranks. It also shows that the learned random variables are typically insufficient to enable the \emph{accuracy-optimal} baseline for arbitrary paths---the sizes of variables with high rank (e.g.,$|V|\geqslant 4$) are small.

\textbf{Histogram Approximation: }
We evaluate the accuracy and space savings of the histogram representations of the learned random variables.
Recall that our method is able to automatically identify the number of buckets per dimension (cf.\ Section~\ref{sssec:repuni}). We call this method \emph{Auto}.
We compare \emph{Auto} with Gaussian~\cite{DBLP:conf/aips/NikolovaBK06}, Gamma~\cite{DBLP:journals/corr/abs-1302-4987}, and exponential~\cite{DBLP:conf/uai/WellmanFL95} distributions.
Figures~\ref{fig:aal-kl} and~\ref{fig:bj-kl} show the KL divergences between the raw cost distribution and the estimated distributions using the different methods.
The results of using exponential distributions are not shown since their KL divergences exceed $1.0$ and are significantly worse than the other ones.
The results clearly suggest that \emph{Auto} provides the most accurate estimation and that travel-time distributions typically do not follow standard distributions.
\emph{Auto} adaptively determines the bucket count for each dimension and then optimally selects the bucket boundaries, thus being able to represent arbitrary distributions.

We compare \emph{Auto} with a static histogram approach that uses a fixed number of buckets per dimension. The method that uses $b$ buckets per dimension is called \emph{Sta}-\textit{b}.
Figure~\ref{fig:MV-opt:accur} shows the KL divergences between the raw distribution that is obtained from the trajectories' travel costs and the different histograms.
As the number of buckets increases, \emph{Sta}-$b$ produces a smaller KL divergence value because the more buckets a histogram has, the better accuracy the histogram can achieve. \emph{Auto} is able to achieve as good accuracy as \emph{Sta}-$4$. This suggests that the \emph{Auto} method is effective.

We evaluate the space-savings achieved by the histogram representation. Intuitively, the more buckets a histogram has, the more storage it needs.
We report the compression ratio $1-\frac{S_H}{S_T}$, where $S_H$ and $S_T$ represent the storage required by the histograms and the underlying raw cost distribution.
The raw cost distribution is of the form $(\mathit{cost}, \mathit{frequency})$. The higher the ratio is, the better the space-savings achieved by the histograms are.
Figure~\ref{fig:MV-opt:saved-space} shows that \emph{Auto} has a better compression ratio than has \emph{Sta}-$4$. This suggests that \emph{Auto} achieves a good trade-off between accuracy and space-saving.

\subsubsection{Accuracy Evaluation}

\textbf{Comparison with Ground Truth: }
We select $100$ paths where each path has more than $\beta=30$ trajectories during half an hour.
For each such path, we record all trajectories that occurred on it during the interval of interest. We use these trajectories to compute the ground-truth distribution using the accuracy-optimal baseline.
Next, we exclude these trajectories from the trajectory data set. Thus, we have the data sparseness problem, and the accuracy-optimal baseline does not work.

We consider the following four methods for estimating the travel cost distributions. (a) \emph{OCRV}: the proposed Optimal Candidate Random Variables based method. (b) \emph{LB}: the legacy baseline that is based on convolution as described in Section~\ref{ssec:aba}. \emph{LB} is regarded as one of the state-of-the-art approaches used in the conventional paradigm~\cite{DBLP:conf/icde/YangGJKS14}. In our setting, \emph{LB} only considers the random variables with ranks one.
(c) \emph{HP}~\cite{DBLP:conf/edbt/HuaP10}: this method assumes that the joint distributions for every pair of edges in a path are known and then computes the joint probability distribution of the path taking these into account. In our setting, this means that HP only considers random variables with ranks two.
(d) \emph{CRV}: this method computes an estimated distribution using a randomly chosen candidate random variable set rather than the optimal set.

First, we consider a concrete example shown in Figure~\ref{fig:introexample}(a). The distributions estimated using \emph{OCRV}, \emph{LB}, \emph{HP} and \emph{CRV} are shown in Figures~\ref{fig:overall-detailed}(a)-(d).
It is clear that \textit{OCRV} captures the main characteristics of the ground-truth distribution. %
The convolution-based estimation \emph{LB} seems to approach a Gaussian distribution (cf.\ the Central Limit Theorem). However, it is clear that a Gaussian distribution is unable to capture the ground-truth distribution, and \emph{LB} is inaccurate.
The distribution computed by \emph{HP} is also inaccurate, which suggests that the dependencies among the edges in a path cannot be fully captured by only considering the dependencies between adjacent edges.
Method \emph{CRV} suggests that a randomly chosen candidate random variable set provides a less accurate estimation compared to the estimation based on the optimal candidate random variable set.

Next, we report results when using paths with different cardinalities. Specifically, we vary $|P_{query}|$ from $5$ to $20$.
Figure~\ref{fig:overall-com} shows the KL-divergence values $\mathit{KL}(p, \hat{p})$, where $p$ is the ground-truth distribution derived by the accuracy-optimal baseline and $\hat{p}$ is the estimated distribution using \emph{OCRV}, \emph{LB}, \emph{CRV}, and \emph{HP}.
As the number of edges in a path increases, the benefits of using the proposed \emph{OCRV} becomes more significant.
In particular, the KL-divergence values of \emph{OCRV} grow slowly while the KL-divergence values of \textit{LB} grow quickly. This is not surprising because \emph{LB} assumes independencies, and the longer a path is, the more likely it is that the edges in the path are not independent.
Next, \emph{OCRV} is also better than \emph{CRV}, which suggests that the optimal candidate random variable set produces the most accurate estimation.
Further, \emph{HP} is better than \emph{LB}, which is because \emph{HP} considers the correlation between adjacent edges. However, \emph{HP} always has larger KL-divergence values than do \emph{CRV} and \emph{OCRV}. This is because coarser random variable sets have smaller KL-divergence (cf.\ Theorem~\ref{th:2}).
In summary, Figure~\ref{fig:overall-com} suggests that the proposed \emph{OCRV} is able to accurately estimate travel cost distributions and that it outperforms the other methods, especially for long pathes.

\textbf{Entropy Comparison: }
Next, we consider long paths where we do not have ground truth distributions. We randomly choose $1,000$ paths for each cardinality with an arbitrary departure time and report average values; and we vary the path cardinality from $20$ to $100$.
Figure~\ref{fig:joint-prob} shows that \emph{OCRV} produces the least entropy, which is consistent with the design of identifying the optimal candidate random variable set. This suggests that the proposed method is able to accurately estimate the distribution of a path.

\subsubsection{Efficiency}

\textbf{Efficiency of Deriving Learned Random Variables: } Since deriving the learned random variables is an off-line task, the run-time is not critical. The procedure can be parallelized in a straightforward manner. Using the default parameter setting, it takes less than $2$ minute with $48$ threads to learn the random variables from $D_1$, and it takes around $45$ minutes with $48$ threads to learn random variables from $D_2$.
This also suggests that when receiving new trajectories regulary, the procedure can be conducted periodically to efficiently update the learned random variables.

\begin{figure}[!htbp]
  \centering
  \subfigure[$D_1$]{
    \label{fig:overall-eva-rt:d1}
    \includegraphics[width=0.23\textwidth]{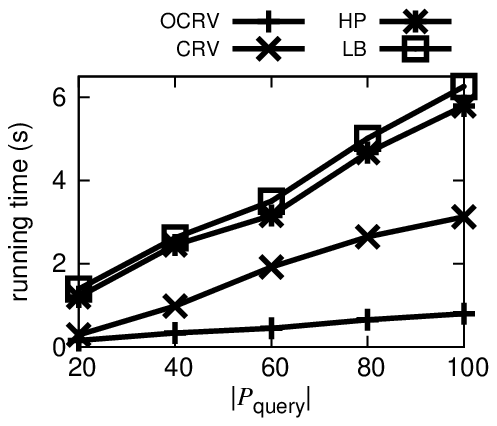}}
  \subfigure[$D_2$]{
    \label{fig:overall-eva-rt:d2}
    \includegraphics[width=0.23\textwidth]{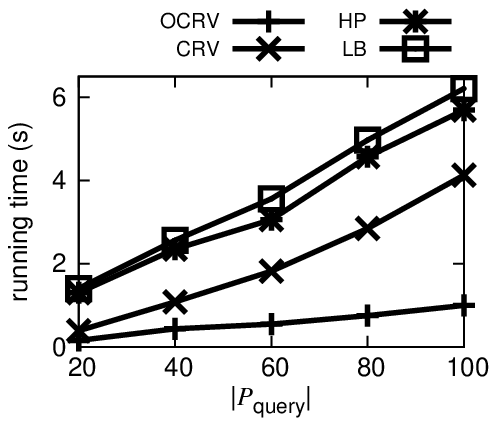}}
  \caption{Efficiency}
  \label{fig:overall-com-ef}
\end{figure}

\textbf{Efficiency of Estimating the Cost Distribution of a Path: } This is an on-line task.
Figure~\ref{fig:overall-com-ef} reports the run-times of different methods. As the cardinality of a query path increases, the run-time also increases. %
As the \emph{HP} method only considers the learned random variables with rank at most two, it has to consider at least $|\mathcal{P}_{\mathit{query}}|$ learned random variables to compute the joint distribution of the path, which makes its running time slightly lower than that of \emph{LB}. In contrast,  \emph{OCRV} and \emph{CRV} are able to exploit learned random variables with higher ranks. Thus, they are able to use significantly fewer learned random variables and are thus faster than \emph{HP}. Since \emph{OCRV} has coarser random variables than does \emph{CRV}, it is able to use fewer learned random variables and is thus faster than \emph{CRV}.
Figure~\ref{fig:overall-com-ef} clearly shows that \emph{OCRV} is the most efficient.

To further investigate the run-time of \emph{OCRV}, a detailed analysis of the run-time of the three major steps used in \emph{OCRV} on paths with cardinality 20 is reported in Figure~\ref{fig:overall-eva:rt}, where differently sized subsets of trajectories are used.
First, the optimal candidate random variable set is identified, denoted by $\emph{OI}$. Thanks to Theorems~\ref{th:3}, this part is very efficient.
Second, the joint distribution is computed according to Equation~\ref{eq:est}, denoted by $\emph{JC}$. This is the most time-consuming part as it needs to go through many hyper-buckets of the histograms in order to compute the joint distributions according to Equation~\ref{eq:est}. However, when having more trajectories, the run-time of $\emph{JC}$ decreases, as we have more learned random variables with higher ranks. Thus, as data volumes increase, performance improves.
Third, deriving the marginal distribution (denoted by $\emph{MC}$) is also very efficient.

\subsubsection{Memory Use}

\begin{figure}[!ht]
   \begin{minipage}[b]{0.24\textwidth}
     \centering
     \includegraphics[width=\columnwidth]{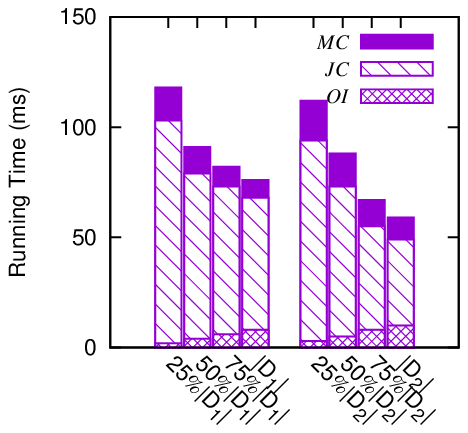}
\caption{Run-Time Analysis}
     \label{fig:overall-eva:rt}
   \end{minipage}
   \hfill
   \begin{minipage}[b]{0.24\textwidth}
     \centering
     \includegraphics[width=\columnwidth]{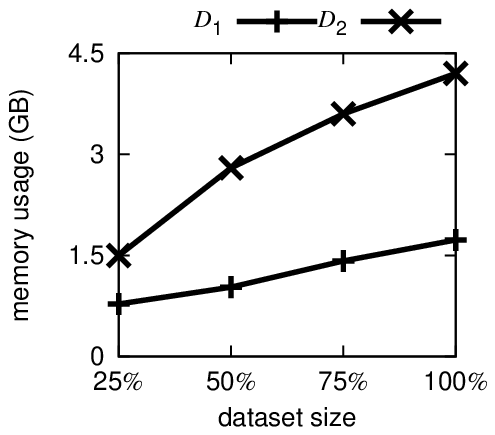}
     \caption{Memory Usage} \label{fig:overall-eva:mem}
   \end{minipage}
\end{figure}

\emph{OCRV} requires memory to store the learned random variables.
As the size of the trajectory data set grows, the memory use also grows, as shown in Figure~\ref{fig:overall-eva:mem}.
Since we use histograms to represent the distributions of learned random variables, the memory use of the \emph{OCRV} is sufficiently small that \emph{OCRV} can be accommodated in main memory. In particular, the learned random variables for Aalborg and Beijiing occupy around 1.8 GB and 4.2 GB, respectively.

\subsection{Summary}
The empirical study shows that:
(1) in realistic settings with sparse data, \emph{OCRV} is the most accurate and efficient method;
(2) the proposed histogram representations are able to approximate arbitrary raw cost distributions well using limited space, making it possible to fit the learned random variables into main memory;
(3) \emph{OCRV} is scalable w.r.t. the path cardinality, meaning that \emph{OCRV} is able to support long paths.

We conclude that the proposed \emph{OCRV} method fully addresses the challenges caused by \emph{data sparseness}, \emph{complex distributions}, and \emph{dependencies} and thus is able to provide accurate travel cost distribution estimation. Further, it is efficient, even for long paths, which makes it appropriate for use in stochastic routing algorithms that require efficient evaluation of the cost distributions of multiple candidate paths.

\section{Related Work}
\label{sec:rw}

We review recent studies on estimating \emph{deterministic} and \emph{uncertain} travel costs, respectively.

\textbf{Estimating Deterministic Travel Costs: }The problem of estimating deterministic travel costs has been studied extensively.
Most such studies focus on accurately estimating the travel costs of individual edges, based on which the travel cost of a path is then computed as the sum of the travel costs of its edges.
Some recent studies employ GPS trajectories that generally cover more edges than does loop detector data. However, in many cases, the available trajectories are unable to cover all edges in a road network.
To address the sparseness of the data, some methods~\cite{DBLP:conf/aaai/ZhengN13,DBLP:conf/aaai/IdeS11,DBLP:journals/tkde/YangKJ14,DBLP:conf/kdd/WangZX14} transfer the travel costs of edges that are covered by trajectories to edges that are not covered by trajectories. In particular, the travel costs of an edge can be transferred to its adjacent edges~\cite{DBLP:conf/aaai/IdeS11,DBLP:conf/aaai/ZhengN13} and to edges that are topologically or geographically similar to the edge~\cite{DBLP:journals/tkde/YangKJ14,DBLP:conf/kdd/WangZX14}. In addition, some proposals consider the temporal context~\cite{DBLP:conf/aaai/ZhengN13,DBLP:journals/tkde/YangKJ14,DBLP:conf/kdd/WangZX14}, i.e., peak vs. off-peak hours, while transferring travel costs.
However, these methods do not support travel cost distributions, and they do not model the dependencies among edges. Therefore, they do not apply to the problem we consider.

When all edges have travel costs, it is possible to estimate the travel cost of any path, i.e., by summing up the travel costs of the edges in the paths~\cite{DBLP:conf/aaai/ZhengN13,DBLP:conf/aaai/IdeS11,DBLP:journals/tkde/YangKJ14}.
However, a recent study~\cite{DBLP:conf/kdd/WangZX14} shows that using the sum of travel costs of edges as the travel cost of a path can be inaccurate because it ignores hard-to-formalize aspects of travel, such as turn costs. Instead, a method is proposed to identify an optimal set of sub-paths that can be concatenated into a path. The path's travel cost is then the sum of travel costs of the sub-paths, which can be obtained from trajectories.
However, this method does not support travel cost distributions, and it assumes independence among sub-paths.

\textbf{Estimating Travel Cost Distributions: } Some studies consider the travel cost uncertainty of a path and model this uncertainty. However, their solutions are based on assumptions that do not apply in our setting. First, some studies assume that the travel cost distribution of each edge follows a standard distribution, e.g., a Gaussian distribution. However, the travel cost distribution of a road segment often follows an arbitrary distribution, as shown in recent studies~\cite{DBLP:conf/icde/YangGJKS14,DBLP:journals/pvldb/0002GJ13} and in Figures~\ref{fig:MV-opt} (a) and (b) in Section~\ref{sec:exp}.
We use multi-dimensional histograms to represent arbitrary distributions.

Second, some studies assume that the travel cost distributions on different edges are independent of each other~\cite{chen2005path,lim2013practical}  or conditionally independent given the arrival times at different edges~\cite{DBLP:conf/icde/YangGJKS14} (i.e., mirroring the LB approach covered in Section~\ref{sec:exp}).
The independence assumption often does not hold as suggested in Section~\ref{sec:pre}, and our approach outperforms LB, as shown in Section~\ref{sec:exp}.
Further, with the exception of one study~\cite{DBLP:conf/icde/YangGJKS14}, all above studies use randomly generated distributions to evaluate their proposals. We use large, real trajectory data sets in empirical evaluations. 

The most advanced method, the HP~\cite{DBLP:conf/edbt/HuaP10} approach covered in Section~\ref{sec:exp}, does not make the independence assumption. The HP approach assumes that the travel costs of pairs of adjacent edges are dependent, but it does not consider dependencies among multiple edges in a path. In contrast, we propose a more generic model that employs joint distributions to fully capture the dependencies among all the edges in a path.
In addition, we identify distributions from real-world trajectory data and support time-varying distributions, while the HP approach employs synthetically generated distributions and does not support time-varying distributions.

Although two recent studies~\cite{DBLP:conf/sigmod/Ma0J14,DBLP:journals/tkde/YuanZXS13} employ histograms to represent travel cost distributions, they only consider travel cost distributions on individual edges, and they assume that the travel cost distributions on edges are independent and cannot capture the dependence among edges.

\section{Conclusion and Future Work}
\label{sec:conclusion}

Accurate estimation of the travel cost distribution of a path is a fundamental functionality in spatial-network related applications. We propose techniques that are able to model joint distributions that capture the travel cost dependencies among sub-paths that form longer paths, which in turn enables accurate travel cost estimation of any path using sparse historical trajectory data.
In particular, we learn a set of random variables that capture the joint distributions of paths that are covered by sufficient trajectories.
We then propose methods capable of selecting an optimal subset of the learned random variables such that their corresponding paths cover a query path. The selected variables enable accurate joint distribution estimation of the query path, and the obtained joint distribution can be transferred into a marginal distribution that captures the distribution of the travel cost of the query path.
Empirical studies in realistic settings offer insight into the design properties of the
proposed solution and suggest that it is effective.

This study provides part of the foundation for a new and promising paradigm where travel costs are associated not only with road-network edges, but with sub-paths. The arguably most pertinent next step is to integrate travel cost distribution estimation into stochastic routing algorithms to enable accurate and efficient routing services. It is also of interest to consider individual drivers' driving behavior to support personalized travel cost distribution estimation.

\appendix

\section{Pseudo Code}
\label{ap:code}

Algorithm~\ref{alg:auto} describes the procedure of automatically identifying the number of buckets for a given set of cost values observed in the qualified trajectories, as discussed in Section~\ref{sssec:repuni}.

\begin{algorithm}[!htp]
\LinesNumbered
\SetKwInOut{Input}{Input}
\SetKwInOut{Output}{Output}
\SetKw{Return}{return}
\caption{{Automatically Identify Bucket Number}}
\label{alg:auto}

\Input{Cost Values in Qualified Trajectories: $\mathit{costs}$}
\Output{Bucket Number: $b$}
Split $\mathit{costs}$ into $f$ equal subsets $cc[1], \ldots, cc[f]$;\\
double $prevEb \leftarrow \infty$; double $currEb \leftarrow \infty$; \\
double $b\leftarrow 0$;\\
\Repeat
{$\mathit{currEb}$ is not significant smaller than $\mathit{prevEb}$}
{
    \tcc{$f$-fold error evaluation}
    $\mathit{prevEb} \leftarrow \mathit{currEb}$;\\
    $b \leftarrow b+1$;\\
    \For{$k = 1 \ldots f$}
    {
        $\mathit{train} \leftarrow$ $\mathit{costs} \setminus cc[k]$; $\mathit{test} \leftarrow$ $cc[k]$; \\
        Generate histogram $H_b^k$ based on $\mathit{train}$ using V-optimal with $b$ buckets; \\
        Compute raw cost distribution $D^k$ based on $\mathit{test}$; \\
        $\mathit{currEb} = \mathit{SE}(H_b^k, D^k);$\\
    }
}
\Return{$b-1$};
\end{algorithm}

Algorithm~\ref{alg:CRVopt} describes the procedure of identifying $\mathit{CRV}_{opt}$, as discussed in Section~\ref{sssec:idenopt}. Note that $\mathbb{V}$ denotes a subset of the learned random variables $\mathit{LRV}$ that are spatially and temporally relevant to path $P$.

\begin{algorithm}[!htb]
\LinesNumbered
\SetKwInOut{Input}{Input}
\SetKwInOut{Output}{Output}
\caption{Identify $\mathit{CRV}_{opt}$}
\label{alg:CRVopt}
\Input{Relevant Learned random variables $\mathbb{V}$}
\Output{$\mathit{CRV}_{opt}$}
$\mathit{CRV}_{opt}\leftarrow\emptyset$; $\mathit{CRVS}\leftarrow\emptyset$;\\
Organize $\mathbb{V}$ into a two dimensional array $\mathit{A}$ as described in Section~\ref{sssec:idenopt};\\
\For{each row in $\mathit{A}$}
{
    Construct a candidate random variable set $D$ based on the random variable with the highest rank in the row;\\
    Boolean $\mathit{addIn} \leftarrow$ \emph{true};\\
    \For{each candidate random variable set $\mathit{CRV}$ in $\mathit{CRVS}$}
    {
        \If{$D$ is coarser than $\mathit{CRV}$}
        {
            Remove $\mathit{CRV}$ from $\mathit{CRVS}$;\\
        }
        \ElseIf{$\mathit{CRV}$ is coarser than $D$}
        {
            $\mathit{addIn} \leftarrow$ \emph{false};\\
            \textbf{break};
        }
    }
    \If{$\mathit{addIn}$}
    {
        Add $D$ into $\mathit{CRVS}$;
    }
}
Compute the entropy for each candidate random variable set in $\mathit{CRVS}$; \\
$\mathit{CRV}_{opt}\leftarrow$ the candidate random variable set with the least entropy;

\Return $\mathit{CRV}_{opt}$;
\end{algorithm}


Algorithm~\ref{alg:joint-marginal} describes the procedure of transferring a joint distribution to a marginal distribution, as discussed in Section~\ref{ssec:marginaldist}.

\begin{algorithm}[!htb]
\LinesNumbered
\SetKwInOut{Input}{Input}
\SetKwInOut{Output}{Output}
\caption{From Joint Distribution to Marginal Distribution}
\label{alg:joint-marginal}
\Input{Joint distribution $\hat{p}_{\mathit{CRV}_{opt}}(\mathbf{C}_{\mathcal{P}})=\{\langle \mathit{hb}_i, \mathit{pr}_i \rangle\}$}
\Output{Marginal distribution $p(V_{\mathcal{P}})=\{\langle \mathit{bu}_k, \mathit{pr}_k \rangle\}$}
Initialize an empty histogram $S\leftarrow \emptyset$;\\
\tcc{Joint distribution to Marginal Distribution}
\For{each hyper-bucket $hb_i$ in $\{\langle \mathit{hb}_i, \mathit{pr}_i \rangle\}$}
{
$S\leftarrow S \cup \langle \mathit{HB2BU}(\mathit{hb}_i), \mathit{pr}_i\rangle$;
}

\tcc{Rearrange Marginal Distribution}
Initialize two empty histograms $H$ and $H'$;\\
$H \leftarrow \{S.\langle bu_1, pr_1\rangle\}$;\\
\For{each $S.\langle bu_i, pr_i\rangle$ pair in $S$}
{
    $H' \leftarrow \emptyset$;\\
    \For{each $H.\langle bu_k, pr_k\rangle$ pair in H}
    {
        $H' \leftarrow H' \cup \mathit{Rearrange}(S.\langle bu_i, pr_i\rangle, H.\langle bu_k, pr_k\rangle)$;
    }
    $H \leftarrow H'$;
}
\Return $H$;
\end{algorithm}

In Algorithm~\ref{alg:joint-marginal}, $S.\langle bu_i, pr_i\rangle$ denotes the $i$-th bucket and probability pair in histogram $S$, and $H.\langle bu_k, pr_k\rangle$ denotes the $k$-th bucket and probability pair in histogram $S$.

\section{Shift-And-Enlarge}
\label{ap:shiftandenlarge}

Before introducing the shift-and-enlarge procedure, we first introduce a few concepts.

Given two paths $\mathcal{P}_i$ and $\mathcal{P}_j$, we define $\mathcal{P}_i \setminus \mathcal{P}_j$ as the sequence of edges that are in $\mathcal{P}_i$ but not in $\mathcal{P}_j$. For instance, if $\mathcal{P}_i=\langle e_1, e_2, e_3\rangle$ and $\mathcal{P}_j=\langle e_3, e_4\rangle$, we have $\mathcal{P}_i \setminus \mathcal{P}_j=\langle e_1, e_2\rangle$.

Next, given a random variable $V_{\mathcal{P}_i}^{I_i}$, we denote the minimum and maximum travel costs of the random variable as $V_{\mathcal{P}_i}^{I_i}.\mathit{min}$ and $V_{\mathcal{P}_i}^{I_j}.\mathit{max}$, respectively. Recall that a random variable is represented as a histogram, and thus the minimum and maximum travel costs are the smallest and the largest values that are represented by the histogram buckets.
Given a time interval $I^{\prime}=[t_s, t_e]$ and a random variable $V_{\mathcal{P}_i}^{I_i}$, we define the shifted-and-enlarged interval as
\[
\mathit{SAE}(I^{\prime}, V_{\mathcal{P}_i}^{I_i})=[I^{\prime}.t_s+V_{\mathcal{P}_i}^{I_i}.\mathit{min}, I^{\prime}.t_e+V_{\mathcal{P}_i}^{I_i}.\mathit{max}].
\]

Based on the above, we can define the updated departure time on $\mathcal{P}_2$. The departure time on $\mathcal{P}_2$ is the original departure time $t$ plus the travel time on path $\mathcal{P}_1 \setminus \mathcal{P}_2$.
Since the travel time of path $\mathcal{P}_1 \setminus \mathcal{P}_2$ is uncertain, the updated departure time on $\mathcal{P}_2$ falls into a shifted-and-enlarged interval $\mathit{SAE}([t, t], V_{\mathcal{P}_1\setminus\mathcal{P}_2}^{I_1})$. If $I_2$ intersects the shifted-and-enlarged interval, it is considered as temporally close to the departure time. The same procedure works for the remaining paths $\mathcal{P}_j$, where $j>2$.

For example, assume that the departure time $t$ is 8:00 and the minimum and maximum travel times in random variable $V_{\mathcal{P}_{1}\setminus\mathcal{P}_2}^{I_{1}}$ are 5 and 10 mins, respectively. The departure time on path  $\mathcal{P}_2$ belongs to [8:05, 8:10]. If $I_2$ intersects [8:05, 8:10] then $I_2$ is relevant.

Formally, given an interval $I_j$ where $j\geqslant 2$, $I_j$ is considered as close to the departure time $t$ if $I_j$ intersects $\mathit{SAE}(I^\prime_{j-1}, V_{\mathcal{P}_{j-1}\setminus \mathcal{P}_{j}}^{I_{j-1}})$, where

\[
  I^\prime_{j-1}= \left\{
  \begin{array}{l}
   [t, t], \text{if }j=2; \\ \\
   \mathit{SAE}(I^\prime_{j-2}, V_{\mathcal{P}_{j-2}\setminus \mathcal{P}_{j-1}}^{I_{j-2}}), \text{if } j>2. \\
  \end{array} \right.
\]

\section{Experimental Results on GHG Emissions}
\label{ap:ghg}

In addition to travel-time based travel costs, we also conducted experiments on GHG emissions based travel costs. In particular, we apply the SIDRA-Running model~\cite{DBLP:conf/gis/GuoM0JK12} on the speeds and accelerations, which can be derived from GPS data, to compute GHG emissions cost values.

We present the evaluation results based on GHG emissions next.

\begin{figure}[!htbp]
  \centering
  \subfigure[$D_1$]{
    \label{fig:ghg-joint-prob:d1}
    \includegraphics[width=0.23\textwidth]{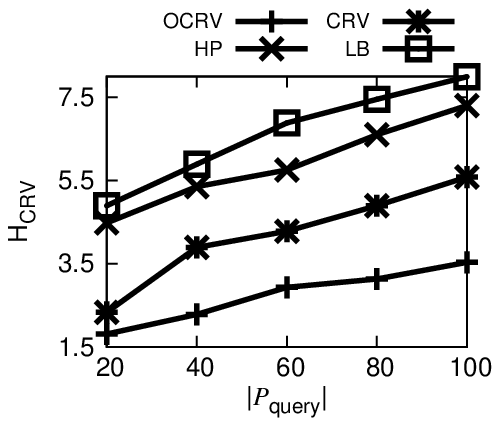}}
  \subfigure[$D_2$]{
    \label{fig:ghg-joint-prob:d2}
    \includegraphics[width=0.23\textwidth]{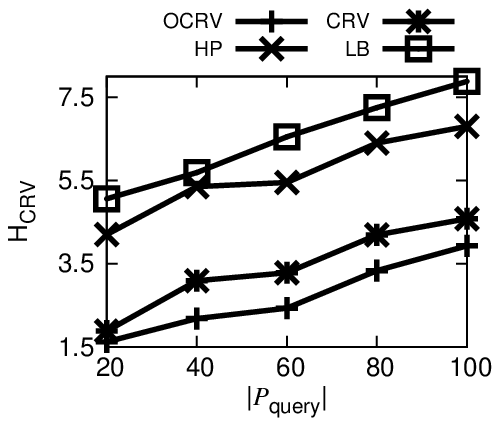}}
  \caption{Entropy Comparison, GHG Emissions}
  \label{fig:ghg-joint-prob}
\end{figure}

\textbf{Entropy Comparison: } Figure~\ref{fig:ghg-joint-prob} shows the entropies among different methods. Similar to the travel-time based results shown in Figure~\ref{fig:joint-prob}, \emph{OCRV} produces the least entropy, suggesting that \emph{OCRV} produces the most accurate estimation.

Compared to Figure~\ref{fig:joint-prob}, the entropy values of GHG emissions distributions are normally smaller than those of travel time distributions. This suggests that diversity of GHG emissions is not as rich as that of travel time.

\begin{figure}[!htbp]
  \centering
  \subfigure[$D_1$]{
    \label{fig:ghg-overall-eva-rt:d1}
    \includegraphics[width=0.23\textwidth]{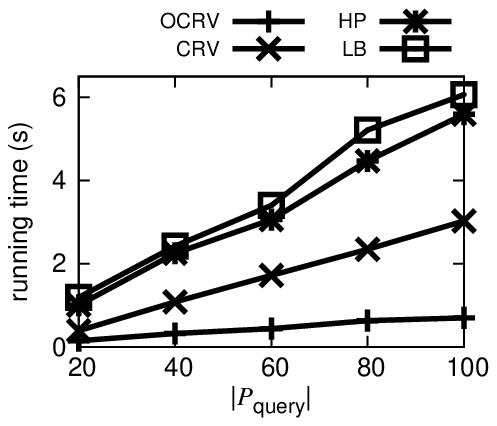}}
  \subfigure[$D_2$]{
    \label{fig:ghg-overall-eva-rt:d2}
    \includegraphics[width=0.23\textwidth]{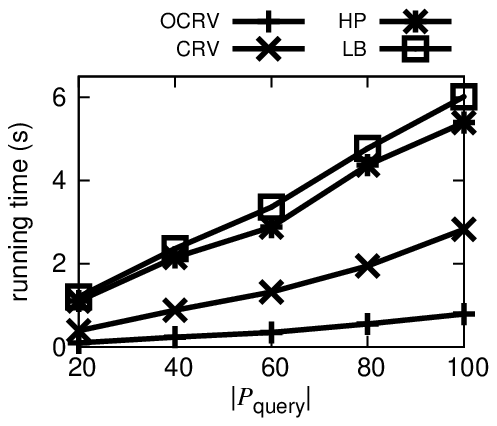}}
  \caption{Efficiency, GHG Emissions}
  \label{fig:ghg-overall-com-ef}
\end{figure}

\textbf{Run-time Comparison: } Figure~\ref{fig:ghg-overall-com-ef} shows that the run-times of the different methods for estimating GHG emissions distributions. As the cardinality of a query path increases, the run-time also increases. Here, \emph{OCRV} outperforms the other methods as it estimates joint distributions using random variables with high ranks. This is consistent with the travel time estimation shown in Figure~\ref{fig:overall-com-ef}.

\begin{figure}[!ht]
   \begin{minipage}[b]{0.24\textwidth}
     \centering
     \includegraphics[width=\columnwidth]{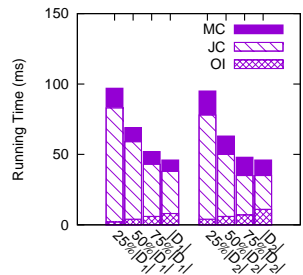}
\caption{Run-Time Analysis}
     \label{fig:ghg-overall-eva:rt}
   \end{minipage}
   \hfill
   \begin{minipage}[b]{0.24\textwidth}
     \centering
     \includegraphics[width=\columnwidth]{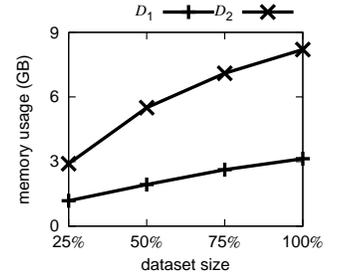}
     \caption{Total Memory Usage} \label{fig:ghg-overall-eva:mem}
   \end{minipage}
\end{figure}

A detailed analysis of the run-time of the three major steps used in \emph{OCRV} on paths with cardinality 20 is reported in Figure~\ref{fig:ghg-overall-eva:rt}. We can see that $\emph{JC}$ also takes the most time and it drops when having more trajectories, as we have more random variables with higher ranks to use. We further notice that the run-time of $\emph{JC}$ for GHG emissions is slightly smaller than the run-time of $\emph{JC}$ for travel time, as depicted in Figure~\ref{fig:ghg-overall-eva:rt}. This is also because that the diversity of GHG emissions is not as rich as that of travel time, which is consistent with the finding we identified in Figure~\ref{fig:ghg-joint-prob}.

\textbf{Memory Use}. We also examine the total memory use for storing the random variables that record both both the GHG emissions and travel time distributions. Figure~\ref{fig:ghg-overall-eva:mem} shows that with the growth of trajectories, the total memory use also grows.
The total memory use is less than twice of the memory use of only travel time based random variables as shown in Figure~\ref{fig:overall-eva:mem}.
This is because both the GHG emissions based random variables and travel time based random variables share the same paths.
Modern servers can easily accommodate the GHG emissions and travel time random variables in main memory and provide efficient in-memory travel costs estimation.

In summary, the proposed \emph{OCRV} method is a generic approach that is able to efficiently and accurately estimate different types of travel cost distributions.


\begin{thebibliography}{10}

\bibitem{ecosky}
C.~Guo, B.~Yang, O.~Andersen, C.~Jensen, and K.~Torp.
\newblock EcoSky: Reducing vehicular environmental impact through eco-routing.
\newblock In {\em {ICDE}}, pages 1412--1415, 2015.

\bibitem{DaiYGD15}
J.~Dai, B.~Yang, C.~Guo and Z.~Ding.
\newblock Personalized route recommendation using big trajectory data.
\newblock In {\em {ICDE}}, pages 543--554, 2015.

\bibitem{bishop2007discrete}
Y.~M. Bishop, S.~E. Fienberg, and P.~W. Holland.
\newblock {\em Discrete multivariate analysis: theory and practice}.
\newblock Springer, 2007.

\bibitem{chen2005path}
A.~Chen and Z.~Ji.
\newblock Path finding under uncertainty.
\newblock {\em Journal of Advanced Transportation} 39(1):19--37, 2005.


\bibitem{cover2012elements}
T.~M. Cover and J.~A. Thomas.
\newblock {\em Elements of information theory}.
\newblock John Wiley \& Sons, 2012.

\bibitem{darroch1983additive}
J.~Darroch and T.~Speed.
\newblock Additive and multiplicative models and interactions.
\newblock {\em Annals of Statistics}, pages 724--738, 1983.

\bibitem{DBLP:conf/gis/GuoM0JK12}
C.~Guo, Y.~Ma, B.~Yang, C.~S. Jensen, and M.~Kaul.
\newblock Ecomark: evaluating models of vehicular environmental impact.
\newblock In {\em{SIGSPATIAL}}, pages 269--278, 2012.

\bibitem{ecomark2}
C.~Guo, B.~Yang, O.~Andersen, C.~Jensen, and K.~Torp.
\newblock Ecomark 2.0: empowering eco-routing with vehicular environmental
  models and actual vehicle fuel consumption data.
\newblock {\em GeoInformatica}, pages 567--599, 2015.

\bibitem{DBLP:conf/edbt/HuaP10}
M.~Hua and J.~Pei.
\newblock Probabilistic path queries in road networks: traffic uncertainty
  aware path selection.
\newblock In {{\em {EDBT}}, pages 347--358, 2010.

\bibitem{DBLP:conf/aaai/IdeS11}
T.~Id{\'{e}} and M.~Sugiyama.
\newblock Trajectory regression on road networks.
\newblock In {{\em AAAI}}, pages 203--208, 2011.

\bibitem{DBLP:conf/vldb/JagadishKMPSS98}
H.~V. Jagadish, N.~Koudas, S.~Muthukrishnan, V.~Poosala, K.~C. Sevcik, and
  T.~Suel.
\newblock Optimal histograms with quality guarantees.
\newblock In {\em {VLDB}}, pages 275--286, 1998.


\bibitem{lim2013practical}
S.~Lim, C.~Sommer, E.~Nikolova, and D.~Rus.
\newblock Practical route planning under delay uncertainty: Stochastic shortest
  path queries.
\newblock {\em Robotics: Science and Systems}, 8:249--256,
  2013.

\bibitem{DBLP:conf/sigmod/Ma0J14}
Y.~Ma, B.~Yang, and C.~S. Jensen.
\newblock Enabling time-dependent uncertain eco-weights for road networks.
\newblock In {\em{GeoRich}}, Article~1, 2014.

\bibitem{malvestuto1991approximating}
F.~M. Malvestuto.
\newblock Approximating discrete probability distributions with decomposable
  models.
\newblock {\em  IEEE Transactions on SMC} 21(5):1287--1294, 1991.

\bibitem{DBLP:conf/gis/NewsonK09}
P.~Newson and J.~Krumm.
\newblock Hidden Markov map matching through noise and sparseness.
\newblock In {\em {SIGSPATIAL}}, pages 336--343, 2009.

\bibitem{DBLP:conf/aips/NikolovaBK06}
E.~Nikolova, M.~Brand, and D.~R. Karger.
\newblock Optimal route planning under uncertainty.
\newblock In {\em{ICAPS}}, pages 131--141, 2006.

\bibitem{smyth2000model}
P.~Smyth.
\newblock Model selection for probabilistic clustering using cross-validated
  likelihood.
\newblock {\em Statistics and Computing} 10(1):63--72, 2000.

\bibitem{DBLP:conf/kdd/WangZX14}
Y.~Wang, Y.~Zheng, and Y.~Xue.
\newblock Travel time estimation of a path using sparse trajectories.
\newblock In {\em {SIGKDD}}, pages 25--34, 2014.

\bibitem{DBLP:conf/uai/WellmanFL95}
M.~P. Wellman, M.~Ford, and K.~Larson.
\newblock Path planning under time-dependent uncertainty.
\newblock In {\em {UAI}}, pages 532--539, 1995.

\bibitem{DBLP:journals/pvldb/0002GJ13}
B.~Yang, C.~Guo, and C.~S. Jensen.
\newblock Travel cost inference from sparse, spatio-temporally correlated time
  series using Markov models.
\newblock In {\em {PVLDB}} 6(9):769--780, 2013.

\bibitem{DBLP:conf/icde/YangGJKS14}
B.~Yang, C.~Guo, C.~S. Jensen, M.~Kaul, and S.~Shang.
\newblock Stochastic skyline route planning under time-varying uncertainty.
\newblock In {\em {ICDE}}, pages 136--147,  2014.

\bibitem{DBLP:journals/tkde/YangKJ14}
B.~Yang, M.~Kaul, and C.~S. Jensen.
\newblock Using incomplete information for complete weight annotation of road
  networks.
\newblock {\em {IEEE} TKDE} 26(5):1267--1279, 2014.

\bibitem{DBLP:journals/tkde/YuanZXS13}
J.~Yuan, Y.~Zheng, X.~Xie, and G.~Sun.
\newblock T-drive: Enhancing driving directions with taxi drivers'
  intelligence.
\newblock {\em {IEEE} TKDE} 25(1):220--232, 2013.

\bibitem{DBLP:journals/corr/abs-1302-4987}
M.P.~Wellman, M.~Ford and K.~Larson.
\newblock Path Planning under Time-Dependent Uncertainty.
\newblock {\em{CoRR}} abs/1302.4987, 2013


\bibitem{DBLP:conf/aaai/ZhengN13}
J.~Zheng and L.~M. Ni.
\newblock Time-dependent trajectory regression on road networks via multi-task
  learning.
\newblock In {\em AAAI}}, pages 1048--1055, 2013.


\end{thebibliography}
\end{document}